\documentclass{emulateapj}

\begin{document}
\title{Probing the Formation of Low Mass X-ray Binaries in Globular Clusters and the Field\altaffilmark{1}}

\author{Arunav Kundu\altaffilmark{2}, Thomas J. Maccarone\altaffilmark{3}, \& Stephen E. Zepf\altaffilmark{2}}

\altaffiltext{1} {Based on observations made with the Chandra X-ray Observatory, which is operated by the Smithsonian Astrophysical Observatory for and on behalf of the National Aeronautics Space Administration under contract NAS8-03060, and the NASA/ESA Hubble Space Telescope which is operated by
    the Association of Universities for Research in Astronomy, Inc., under NASA contract
    NAS 5-26555.}

\altaffiltext{2}{ Physics \& Astronomy Department, Michigan State University, East Lansing, MI 48824; e-mail: akundu, zepf @pa.msu.edu}
\altaffiltext{3}{School of Physics \& Astronomy, University of Southampton, Southampton, Hampshire SO17 1BJ; e-mail: tjm@astro.soton.ac.uk }

\begin{abstract}
 	We present an analysis of low mass X-ray binaries (LMXBs) and globular clusters (GCs) in five early-type galaxies using Chandra X-ray, and HST optical data. Of the 186 LMXBs within the optical fields-of-view 71 are in GCs, confirming that LMXBs are formed particularly efficiently in clusters due to dynamical interactions. However, there is no statistically significant correlation between the distance of a cluster from the center of its host galaxy and its LMXB hosting probability. LMXBs are preferentially found in luminous and metal-rich GCs.  Metal-rich clusters are 3.4 times more likely to host LMXBs than metal-poor ones. This is slightly higher than that measured in other surveys, likely because of larger contamination of the GC sample in previous ground-based datasets, and the inclusion of galaxies with intermediate-age clusters in others. Intriguingly, the LMXBs in NGC 1399 are preferentially in the  reddest clusters of the metal-rich GC subsystem. This indicates that the red peak of the bimodal GC color distribution itself encompasses clusters with a range of enrichment histories. The strength of this effect varies from galaxy to galaxy, possibly indicating differences in their metal-enrichment histories. Field LMXBs in our program galaxies are more concentrated towards the center of their host galaxies than GC-LMXBs. This suggests that a majority of field LMXBs are formed in situ and are not a population that has escaped from current GCs. This is consistent with previous specific frequency based studies. The brightest X-ray sources in GCs appear to be preferentially associated with luminous, metal-rich clusters. We show that it is probable that some of these clusters host multiple bright LMXBs, while the probability is much lower for metal-poor GCs.  If this interpretation is correct, our study implies that LMXBs in more metal-rich cluster systems should reveal a longer high luminosity X-ray tail, and show less X-ray variability than metal-poor cluster populations.

\end{abstract}


\section{Introduction}
	
Chandra images of nearby galaxies reveal large numbers of bright X-ray point sources (Sarazin, Irwin, \& Bregman 2000; Kraft et al. 2001), confirming a long-standing suggestion that the X-ray emission in X-ray faint ellipticals is predominantly from unresolved X-ray binaries (Trinchieri \& Fabbiano 1985).   In elliptical and S0 galaxies most of the bright, L$_X$$\gtrsim$10$^{37}$ erg s$^{-1}$ sources seen in typical Chandra observations must be low mass X-ray binaries (LMXBs) since they generally have stellar populations that are at least a few Gyrs old (e.g. Trager et al. 2000). Chandra observations of these galaxies provide a unique opportunity to probe statistically significant samples of such X-ray bright binary systems consisting of a neutron star or black hole accreting from a low mass companion.

An important characteristic of LMXBs is that they are disproportionately abundant in globular clusters (GC). This has long been attributed to efficient formation of LMXBs in clusters due to dynamical interactions in the core (Clark 1975; Fabian et al. 1975; Hills 1976; Verbunt 1987). In the Milky Way the number of bright LMXBs per unit stellar mass is two orders of magnitude higher in GCs than it is in the field. Such a preference for globular clusters has also been observed in ellipticals and S0s (e.g. Angelini, Loewenstein, \& Mushotzky 2001; Kundu, Maccarone \& Zepf 2002, hereafter KMZ; Sarazin et al. 2003). The identification of LMXBs with globular clusters, which are simple stellar systems with well defined properties such as metallicity and age, provides a unique opportunity to probe the effects of these parameters on LMXB formation and evolution.

Early type galaxies are ideal for studies of the LMXB-GC link as they are particularly abundant in GCs (e.g. Ashman \& Zepf 1998), and are largely unaffected by contamination from high mass X-ray binary systems associated with young stellar populations that complicate the study of spiral galaxies such as the Milky Way. Much progress has been made with recent Chandra studies of LMXBs in ellipticals and S0s. For example roughly half of these binaries are in GCs, and  metal-rich clusters are more likely to host LMXBs (e.g. KMZ, Kim et al. 2006; Verbunt \& Lewin 2006; Sivakoff, Sarazin \& Irwin 2003). 

In this paper we undertake a systematic survey to address some key questions about LMXB formation and evolution. One of these is the significance of the metallicity dependence of LMXBs in GCs. While some studies suggest a strong trend (KMZ; Kundu et al. 2003; Jordan et al. 2004; Kim et al. 2006) other analyses suggest a much weaker effect (Sarazin et al. 2003; Angelini et al. 2001; Di Stefano et al. 2002, 2003). One reason for the discrepancy may be because the ground based data used for one or more of the program galaxies in the latter studies have significantly more contamination in the GC samples than the HST observations studied in this paper. Moreover, some of the galaxies analyzed in those studies are known to host intermediate age GCs (e.g. Kundu et al. 2005 and references therein). The optical colors of such intermediate age GCs are known to be affected significantly by both the age and metallicity of the system, making it harder to quantify the metallicity effect from optical data alone (Kundu et al. 2003). In this paper we present a consistent HST and Chandra analysis of 5 galaxies with known bimodal metallicity distributions to probe the effect of GC metallicity on the probability of hosting a LMXB. We show that metal-rich GCs are 3.4 times as likely to host LMXBs as metal-poor ones in $\S$3.1 and probe the implications of this result in $\S$3.2. We also show in $\S$3.1 that the enhanced LMXB rate in metal-rich GCs is indeed an independent effect and is not a proxy for the galactocentric distance of a GC, which may affect the dynamical evolution rate of a cluster.

A second key question is about the role of field versus GC formation scenarios and what the spatial distributions of field and GC LMXBs tell us about the ancestry of field LMXBs e.g. are they formed in the field or are they an escaped GC population?  Previous studies have yielded conflicting results (Maccarone, Kundu, \& Zepf 2003, hereafter MKZ; Kim et al. 2006). We show in $\S$3.3 that a majority of field LMXBs are unlikely to be ejected from GCs, and more likley to have been formed in situ. We also show that our conclusions are consistent with the dependence of the fraction of LMXBs in GCs with galaxy type  (MKZ; Irwin 2005; Juett 2005). In $\S$3.4 we present a comparison of the fractions of LMXBs in GCs and luminosity functions in various galaxies.

We investigate possible correlations between the X-ray luminosity and GC properties in $\S$3.5. We show that the brightest LMXBs in metal-rich GCs which have been identified as probable black hole candidates may in fact host multiple bright LMXBs. This may explain why a recent study by Irwin (2006) finds that such LMXBs are remarkably stable over time. Finally we present the X-ray, optical and matching source lists for all the galaxies in our sample.

\section{Observations \& Data Reduction}

We  have analyzed archival {\it Chandra} ACIS-S3 images of 5 early type 
galaxies, NGC 1399, NGC 3115, NGC 3379, NGC 4594 \& NGC 4649. The observational details of the Chandra X-ray and HST optical data sets are listed in Table 1. 
The Chandra data were analyzed and point sources likely to be associated with LMXBs identified according to the procedure laid out in KMZ and MKZ.  In brief, after standard pipeline processing of the Chandra data \footnote {http://www.astro.psu.edu/xray/acis/recipes/clean.html} we identified X-ray point sources in the 0.5-2.0, 2.0-8.0, and 0.5-8 keV images using WAVDETECT in the CIAO package with a threshold of 10$^{-6}$ probability of false
detections ($\lesssim$1 false source per field). We consider all sources within an arcsec of each other to be multiple detections of the same object. Further visual examination suggests that a handful of candidates that had a larger separation were also likely to be duplicate observations with centering errors in one or more of the energy bands. We retained the coordinates of the object in the band with most counts (and hence the best determined center) in these cases.
For each LMXB we attempted to fit the spectral index and calculate the flux. For cases where the spectral index could not be fit due to the lack of photons, we fixed the spectral index to $\Gamma$=1.7, where $\frac{dN}{dE}$=E$^{-Gamma}$. We also fixed the spectral indices of all the NGC 1399 sources, where the hot gas made background subtraction difficult. The luminosities of LMXBs were calculated using the Tonry et al. (2001) surface brightness fluctuation distances to each galaxy.

Some aspects of the LMXB properties in several of these galaxies have been analyzed in previous studies (Angelini et al. 2001; Di Stefano et al. 2003; Sarazin et al. 2003; Randall et al. 2004; Kim et al. 2006). The X-ray properties of our sample are consistent with these studies in the areas that these analyses overlap. However, the some of the correlations between various GC and LMXB properties do differ between our study and some of these previous ones. We comment on these differences in the relevant sections.

	The  globular cluster system in the inner regions of the program galaxies have previously been studied using HST-WFPC2 by us (Kundu \& Whitmore 1998, 2001) and other groups (e.g. Grillmair et al. 1999; Larsen et al. 2001). 
These five galaxies have specifically been chosen because they have clearly bimodal cluster colors, with minimal background contamination, in order to probe the effect of metallicity on LMXB formation. KMM mixture modelling tests (Ashman, Bird, \& Zepf 1994) reveal that the color distribution of clusters in each of these galaxies is better fit by a bimodal distribution than a unimodal one at better than 95\% confidence (Kundu \& Whitmore 1998, 2001). We note that 
although the NGC 3379 is bimodal according to this test we classified it as 'likely' bimodal in Kundu \& Whitmore (2001) because of its relatively sparse globular cluster system, and the recommendation of abundant caution in interpreting the KMM results of small samples by Ashman et al. (1994). We include this galaxy in our sample both because it illustrates the effects of small number statistics on the interpretation of the GC-LMXB connection and because it is the subject of a very deep Chandra imaging survey (Fabbiano et al. 2006) and the optical data we present here will be useful for future analysis of this galaxy.

 Studies of globular cluster systems agree that bimodal GC distributions primarily reflect differences in the metallicities in two sub-populations of old ($\gtrsim$8 Gyrs) GCs (e.g. Puzia et al. 2002; Cohen, Blakeslee \& Cote 2003; Hempel \& Kissler-Patig 2004; Kundu et al. 2005), while unimodal color distributions may indicate an age spread (see Kundu et al. 2005 and references within). Hence, the galaxies in our sample have bimodal metallicity distributions. The blue peak corresponds to the metal-poor GCs, while the red sub-population is associated with the metal-rich clusters.

The observational details of the WFPC2 data are listed in Table 1. For this paper we use the globular cluster lists of NGC 3115, NGC 3379, NGC 4649 \& NGC 1399 from Kundu \& Whitmore (1998, 2001) and Kundu et al. (2005). For consistency we have reanalyzed the cluster systems of NGC 1399 and NGC 4594 using the procedure outlined in Kundu \& Whitmore (2001) and references therein. We compared these results with the published studies of Grillmair et al. (1999) and Larsen et al. (2001) and found good photometric agreement. However, the Grillmair et al. (1999) data applied a very broad color cut that allowed contaminating sources in their GC lists. The effects of this on previous GC-LMXB studies is discussed below. We note that while we used the Holtzman et al. (1995) on-orbit calibration to convert the WFPC2 F555W and F814W measurements of NGC 3115, NGC 3379 and NGC 4649  to V and I, the pre-launch synthetic calibration of F547M and F814W (Holtzman et al. 1995) is the only one available  to convert the NGC 4594 data to V and I. The F450W and F814W observations of NGC 1399 were calibrated to B and I using on-orbit calibrations generously provided to us by Jon Holtzman (private communication).
We note that while we have measured the half-light radii of the globular clusters in these galaxies (e.g. Kundu \& Whitmore 2001), we do not study the effect of this parameter on the LMXB formation rate in GCs because it is the core radius that is crucial for understanding the interaction rate in the cores of GCs where LMXBs are typically found. The small core radii of GCs at these distances can only be constrained with extremely deep HST data, and the apparent constraints obtained from shallower data are unreliable. For example we showed in Smits et al. (2006) that including the core radii of M87 globular clusters estimated by Jordan et al. (2004) did not improve the predictive power of the most likely relation for whether an X-ray binary will exist in a globular cluster, compared to what is learned by just considering the effect of mass.  The SBF distances from Tonry et al. (2001) have been adopted throughout this paper when calculating distance dependent quantities. Burstein \& Heiles (1982) foreground reddening corrections have been adopted for all galaxies for consistency with earlier GC analysis.

\subsection{The Sample}

\subsubsection{NGC 1399 }

	NGC 1399, the giant elliptical galaxy at the center of the Fornax cluster, has a rich and well studied globular cluster system (e.g. Grillmair et al. 1999; Dirsch et al. 2003). Our resulting sample is composed of 175 bona fide LMXB candidates (Table 2). About 10-15 of these objects are likely to be associated with contaminants such as background AGNs (Brandt et al. 2000; Mushotzky et al. 2000); the rest are expected to be LMXBs.   

	Our analysis of the NGC 1399 globular cluster system using the method outlined in Kundu et al. (2005) and references therein identified 554 candidates with colors between 1.5$<$(B-I)$<$2.5. We note that only objects in this color range are considered to be bona fide cluster candidates as stellar evolutionary models (e.g. Bruzual \& Charlot 2003; Maraston 2005) indicate that this encompasses the full spread of possible GC colors i.e. all colors for a zero redshift stellar population. That all GCs fall well within this color range is further supported both by Galactic GC data and by spectroscopic follow-up of extragalactic GC systems (Sharples et al. 1998; Zepf et al. 2000). The coordinate system of the X-ray and optical images were bootstrapped using obvious LMXB GC matches according to the technique outlined in MKZ. As in our previous analyses we achieve a relative astrometric accuracy of better that 0.3 arcsecs (for all galaxies in this paper).  The first few candidates in the optical source list are published in Table 3 and the complete list in electronic format. All lists in this paper are sorted by distance from the center of the respective galaxy.

 Thirty eight of the LMXB candidates in the ACIS data lie within the WFPC2 field of view (not including 4 objects that are not considered bona fide LMXBs for the reasons outlined below). A substantial fraction of the identified LMXB population is thus in the WFPC2 field even though it is much smaller than the ACIS field. This is because most LMXBs are found in the main body of the galaxy. NGC 1399 is a X-ray bright galaxy with a significant component of hot gas. However, most of the emission is concentrated in the inner regions of the galaxy. Visual inspection of the ACIS data reveals that the LMXB candidates outside a radius of 8 arcsecs from the center are secure detections that can be used for further analysis of the GC-LMXB link. All but the nuclear source lie outside this region. This is not to say that the X-ray completeness is uniform outside of 8'', or that there are no X-ray binaries in the regions with X-ray bright, hot gas. We exclude the innermost regions (in this and other X-ray bright galaxies in our sample) in order to study the most reliable sample of LMXBs. Where appropriate we comment on the possible effect of incompleteness on our analysis. For consistency we only consider the 548 GC candidates that lie outside the inner 8 arcsecs for LMXB-GC studies.  X-ray sources 11, 24 and 40 (see Table 2) are associated with optical candidates bluer than B-I=1.5 and are likely associated with contaminants (e.g. AGNs). They are excluded from the list of bona fide LMXBs within the WFPC2 frame. 

Twenty four of the thirty eight LMXBs within the WFPC2 field of view are within 0.6 arcsecs of a globular cluster candidate and considered to be GC-LMXB sources. The expected number of false matches at this matching radius is 1.5 objects. Increasing the matching radius to 0.7 arcsecs does not add any new sources. Relaxing the radius to 1 arcsec yields 6 new candidates with a corresponding increase of the expected false matches to 4.2. Given the likelihood that most of these extra
matches are spurious and may skew statistical tests for correlations between GC and LMXB properties we choose to adopt the tighter constraint (for this and other galaxies in the study). The LMXB-GC candidates are presented in Table 4. The tables with complete lists of the GC and LMXB candidates in each of our candidate galaxies is available in electronic format.

\subsubsection{NGC 3115 }

We presented a partial analysis of the GC-LMXBs in NGC 3115, the nearby bulge dominated S0 galaxy, in Kundu et al. (2003) in the context of that study. In this paper we present a detailed analysis along with the data. We detect 90 LMXB candidates that are separated by at least an arcsec, although a closer visual inspection of the ACIS-S3 chip  reveals that 4 of the candidates near the edge of the image are likely to be multiple detections of the same object in different bands due to the extended nature of the off-axis point spread function. Of the 86 bona fide LMXB candidates 36 (excluding the central AGN) fall within the HST-WFPC2 field of view. 

	We use the globular cluster data from the Kundu \& Whitmore (1998) analysis of the globular cluster system of NGC 3115 within the HST field of view, and refer the reader to that study for details of the globular cluster system. There are 133 GC candidates with colors between 
0.8$<$(V-I)$<$1.4 in the WFPC2 image. Of these nine GCs are within 0.5 arcsecs of a LMXB and are considered GC-LMXBs. Expanding the matching radius to 1 arcsec does not add any further candidates. The expected number of false matches for a matching radius of 0.5 arcsecs is only 0.2 sources. The source lists are published in Tables 5 \& 6 and the LMXB-GC candidates in Table 7.

\subsubsection{NGC 3379 }

NGC 3379 is an X-ray faint elliptical in the nearby Leo Group of galaxies with 
a relatively sparse globular cluster system (Kundu \& Whitmore 2001). We identify 70 candidate LMXBs in the ACIS-S3 field of view. Twenty six of these sources lie within the WFPC2 field of view. This does not include X-ray source 19 that lies on the edge of the WFPC2 chip and is possibly associated with  an optical counterpart that lies in the vignetted region of the chip. Since this X-ray source cannot reliably be associated with either a GC or the field it is not considered for further statistical analysis.

The globular cluster study of Kundu \& Whitmore (2001) identified 61 globular clusters with colors between 0.8$<$(V-I)$<$1.4 in the WFPC2 field. Of these 7 GCs are within 0.5 arcsecs of an LMXB candidate and are considered to be GC-LMXB matches. No further candidates are added even when the matching radius is increased to 1 arcsec. Only 0.1 false matches are expected for a random distribution of GCs and LMXBs. The X-ray, optical and GC-LMXB lists are published in in Tables 8, 9 \& 10 respectively.

\subsubsection{NGC 4594 }

 NGC 4594 (The Sombrero) is a nearby bulge dominated galaxy with a thin edge on disk that is formally classified as a Sa galaxy, although it has a bulge to disk ratio of 6 (e.g. Kent 1988). We detect 141 LMXB candidates in the 0.5-8 keV energy range in the ACIS-S3 
observations of NGC 4594. Forty nine of these candidates lie within the WFPC2 field of view, excluding the source associated with the nucleus which is likely linked to emission from the central black hole. Another two sources (X-ray IDs 14 and 65) are associated with objects that are redder than typical GCs. These happen to fall in dust lanes and it is possible that they are reddened by dust, rather than being background objects. However, as these candidates may also be associated with stellar populations and it is not clear which population they belong to we do not consider them in either the GC or the field LMXB lists for statistical tests.  

 We detect 193 GC candidates with a clearly bimodal distribution of colors between 0.8$<$(V-I)$<$1.4 in the WFPC2 image of the central region. Of these 15 are LMXB-GC sources that lie within 0.6 arcsecs of an LMXB with 0.7 likely false matches for a random distribution of objects.  The LMXB, GC and GC-LMXB object lists are published in Tables 11, 12 \& 13 respectively.

\subsubsection {NGC 4649}

	NGC 4649 is an X-ray bright elliptical in the Virgo cluster with a rich and clearly bimodal globular cluster system (Kundu \& Whitmore 2001). 	We identify 165 point-like X-ray sources in the ACIS-S3 image of NGC 4649 (Table 14). Candidate 1 in the X-ray list is the nuclear source associated with the central black hole. The optical counterpart of source 2 is bluer than a typical GC, while source 31 is too red. NGC 4649 also has a nearby spiral companion NGC 4647. The edge of this galaxy is seen in the WFPC2 image of NGC 4649. In order to reduce the possibility of contamination by high mass X-ray binaries associated with this galaxy we do not consider the X-ray sources 71, 73, 74 and 87 that lie near NGC 4647 for the rest of the analysis. After removing these likely interlopers there are 37 LMXBs within the HST-WFPC2 field of view. We note that the X-ray bright, hot gas is NGC 4649 is concentrated towards the center of the galaxy. The 37 LMXBs considered here are all farther than 10 arcsecs from the nucleus, where the background is largely resolved by Chandra. 

	The Kundu \& Whitmore (2001) globular cluster study of NGC 4649 identified 418 candidates with colors between 0.8$<$(V-I)$<$1.4 in the WFPC2
 image. After eliminating 13 sources that lie within 10 arcsecs of the center of the galaxy and hence coincident with the hot gas, there are 405 GC candidates in the rest of the optical image. We note that none of the GCs are obviously associated with the companion spiral NGC 4647. Of the 405 GCs under consideration, 16 lie within 0.6 arcsecs of a LMXB and are considered to GC-LMXB matches with 1 likely false match for a random distribution of sources.  Tables for the LMXB, GC and GC-LMXB object lists are published in Tables 14, 15 \& 16 respectively.

\section {Analysis \& Discussion}

\subsection{Are Metal-Rich Globular Clusters preferential LMXB hosts? }

\begin{figure*}
\includegraphics[angle=-90, scale=0.7]{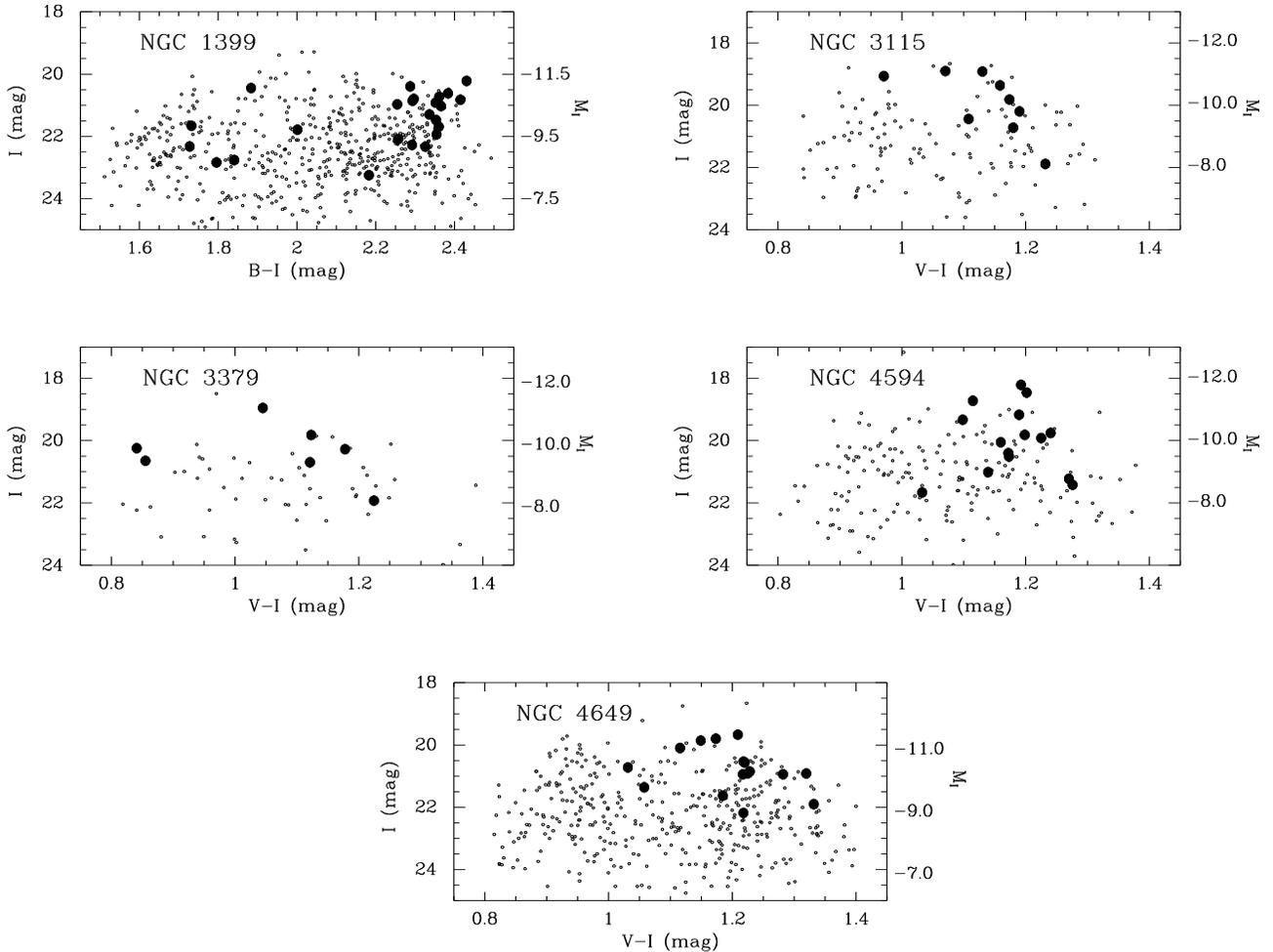}
\caption{ Color-magnitude diagrams for globular cluster candidates in the 5 program galaxies. Filled points represent clusters with LMXB counterparts. LMXBs are clearly found preferentially in luminous (high mass) and red (metal-rich) globular clusters. The absolute magnitude scale for each galaxy, using the distance modulus from Table 1, is shown on the right side of each plot. We note that the turnover magnitude of the Gaussian globular cluster luminosity function is at an absolute magnitude of M$_I$$\approx$8.5 (Kundu \& Whitmore 2001).  }
\end{figure*}

	A surprising conclusion of our study of the LMXBs in NGC 4472 (KMZ) is that red, metal-rich GCs are about 3 times more likely to host LMXBs than blue, metal-poor ones. While a similar effect is seen in the Milky Way, the small number of LMXBs in GCs (thirteen bright LMXBs) and the preferred location of the handful of galactic metal-rich GCs in the bulge makes it difficult to disentangle metallicity effects from enhanced dynamical evolution on GCs close to the Galactic center (e.g. Grindlay 1987; Bellazzini 1995).  Some recent studies of ellipticals and S0s appear to confirm the strong trend with metallicity seen in NGC 4472 (Kundu et al. 2003; Jordan et al. 2004; Kim et al. 2006). However, others suggest a weaker effect (Sarazin et al. 2003; Angelini et al. 2001; Di Stefano et al. 2003). These variations might be due to limited sample size, contamination by background sources such as AGN, and/or the inclusion of  galaxies with unimodal color distribution that may reflect a population of younger clusters in some of the data sets. The confirmed bimodality in our GC samples, negligible contamination in the WFPC2 GC sample (Kundu \& Whitmore 2001), and the range of sample sizes in or program galaxies allows us to control and test for these effects. 

\subsubsection{Colors and Magnitudes of GCs hosting LMXBs}

 The color-magnitude diagrams of globular clusters in the 5 program galaxies are presented in Fig 1, with filled points representing the LMXB hosts. The bimodal color distribution of GCs is apparent. Figure 1 reveals that LMXBs are found preferentially in more massive and red (metal-rich) GCs  confirming the trend seen in KMZ. These properties can be seen more clearly in Fig 2 where we plot histograms of the color and luminosity distributions of the GCs in each of the galaxies along with those of the LMXB hosts. The LMXB hosts are clearly associated with the reddest and brightest globular clusters. 

The median I magnitudes of the globular clusters in the WFPC2 field and of the subset hosting LMXBs in each of the program galaxies are: 22.42:21.39 for NGC 1399,  21.24:19.82 for NGC 3115,  21.45:20.28 for NGC 3379,  21.13:19.92 for NGC 4594, and  22.24:20.89 for NGC 4649 respectively. These data clearly show that LMXBs preferentially populate bright GCs. 

Similarly the median V-I colors of all GCs in the HST field of view (B-I for NGC 1399) and GCs with LMXBs in our galaxies are: 2.08:2.29 for NGC 1399,  1.07:1.16 for NGC 3115,  1.10:1.12 for NGC 3379,  1.09:1.19 for NGC 4594, and  1.11:1.21 for NGC 4649 respectively, revealing that redder, metal-rich GCs are preferred hosts of bright LMXBs. Not surprisingly the richest globular cluster systems such as NGC 1399 and NGC 4649 show the most convincing evidence of the color effect while the effect is less obvious in the sparse GC (and LMXB) system of NGC 3379. Thus it is unlikely that even multi-epoch, deep, Chandra X-ray data sets, such as the one currently beng obtained (Fabbiano et al. 2006), will yield a statistically strong metallicity signature in this galaxy.

\begin{figure}
\includegraphics[angle=0, scale=0.55]{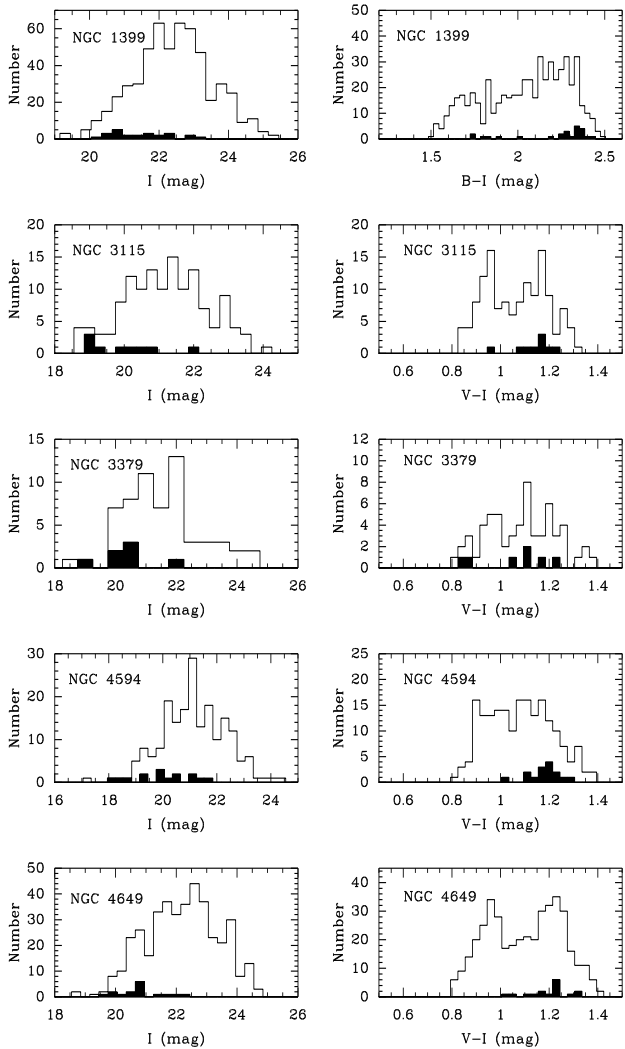}
\caption{ Histograms of the luminosity (left) and color (right) distributions of globular clusters in each of the program galaxies. The filled histograms represent the subset of clusters that are LMXB hosts. The left panels show that bright, and hence more massive, globular clusters are preferred hosts for LMXBs. The bimodal nature of the globular cluster color (metallicity) distribution can clearly be seen in the right panels. Metal-rich GCs are more than three times as likely to host bright LMXBs as metal-poor GCs. }
\end{figure}

Next we divide the GCs in each galaxy into metal-rich and metal-poor subsystems using the KMM test (Ashman et al. 1994) and calculate the fraction of LMXBs in the two subsystems. The fraction of LMXBs in the blue and red subsystems in our program galaxies are 2.8\%:5.6\% in NGC 1399, 1.8\%:13.8\% in NGC 3115, 13.6\%:13.8\% in NGC 3379, 1.3\%:15.7\% in NGC 4594, and 1.2\% and 13.8\% in NGC 4649. We note that in order to reduce the effects of color selection bias at the faint end of GC distribution, which depends on the choice of filter and the relative depths of the optical data in the two filters, we only considered GCs that are brighter than 1 mag past the M$_I$$\approx$-8.46 mags globular cluster luminosity function turnover (Kundu \& Whitmore 2001) in each galaxy. Since LMXBs are preferentially located in bright GCs an excess of faint GCs at either end of the color distribution can skew the red vs. blue statistics if this criteria is not applied. The combined sample of 71 GC-LMXBs, and corresponding GCs that are brighter than a magnitude past the turnover in our 5 galaxies  yields a ratio of 1:3.4 for LMXBs in blue and red GCs, in excellent agreement with the 1:3.3 ratio in NGC 4472 (KMZ). We note that although this is consistent with the Kim et al. (2006) observations, it is larger than the 1:2.8 ratio derived by them. This is likely due to the larger contamination in the GC sample of the ground based data analyzed by Kim et al. (2006). In all galaxies studied to date the metal-poor GC sub-system has a larger core radius than the metal-rich population (e.g. Geisler, Lee, \& Kim 1996) and hence the predominantly blue candidate GCs and GC-LMXBs in the outer halo are likely to be preferentially contaminated.

The color distributions of GC-LMXBs of some of these galaxies has been analyzed in previous studies, which reach somewhat different results.  Angelini et al. (2001) have studied the GC-LMXB link in NGC 1399. However, they conclude that there is only a ``statistically marginal tendency''  for the GCs containing X-ray sources to be redder than GCs in general. Our data points to a strong metallicity effect in NGC 1399. This discrepancy can likely be traced to the fact that Angelini et al. (2001) do not apply any color cuts to the optical distribution; hence they include contaminating background objects in their analysis. We note that in $\S$2.1.1 we found 3 optical sources with X-ray matches that have colors that are too blue for globular clusters and hence likely to be contaminants. Moreover, the combination of filter choice and the depth of the HST data causes a selection bias towards red GCs at the fainter end which likely affects the Angelini et al. (2001) conclusion. In their analysis of NGC 4594 Di Stefano et al. (2003) conclude that LMXBs are preferentially located in red, metal-rich GCs but find that surprisingly the brightest LMXBs do not show any preference for red GCs and are equally likely to be associated with metal-poor candidates. We find only one GC that is marginally in the blue population of NGC 4594. Given the higher contamination rate in the ground-based GC sample of Di Stefano et al. (2003) it is possible that some of the ``blue" GC-LMXB candidates in their sample are contaminating background objects such as AGN (also see $\S$3.5 below).

\begin{figure}
\includegraphics[angle=0, scale=0.5]{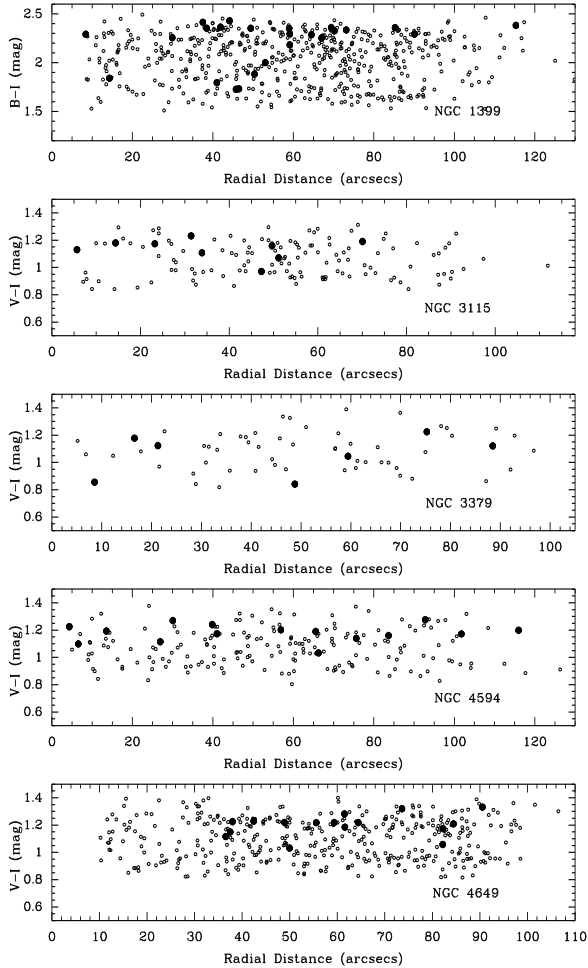}
\caption{ The V-I (B-I for NGC 1399) colors of globular clusters as a function of radial distance from the center of the galaxy. Filled circles represent LMXB hosts. The is no apparent radial trend in the efficiency of formation of bright LMXBs in globular clusters.}
\end{figure}

	 A significant cause of the lack of a strong metallicity effect in the Sarazin et al. (2003) study can be the inclusion of NGC 4365 in the sample. NGC 4365 is one of the few elliptical galaxies that does not have a bimodal optical GC color distribution (e.g. Gebhardt \& Kissler-Patig 1999; Kundu \& Whitmore 2001). Infrared and spectroscopic studies suggest that NGC 4365 has a significant fraction of intermediate color clusters, which are likely formed in a significant star formation 2-8 Gyrs ago (Puzia et al. 2002; Larsen et al. 2006; Kundu et al. 2005 and references therein). Studies of the age and metallicity distributions of cluster systems agree that bimodal GC color distributions 
primarily reflect differences in the metallicities in the two subpopulations of old ($\gtrsim$ 8 Gyrs) clusters (Puzia et al 2002; Cohen et al. 2003, Hempel \& Kissler-Patig 2004; Kundu et al. 2005). The contribution of metal-rich, but relatively blue, intermediate age GCs and GC-LMXBs in NGC 4365 likely skews the statistics based on broad band optical colors in this galaxy (Kundu et al. 2003) and hence the Sarazin et al. (2003) sample.

\subsubsection{Absence of a Galactocentric Distance Effect}

	While our data convincingly show a correlation between the LMXB hosting frequency and the GC metallicity,  studies of globular cluster systems have consistently shown that the metal-rich GC subpopulations are more centrally concentrated than metal-poor clusters (e.g. Geisler at al. 1996), leaving open the possibility that galactocentric distance is the true independent variable.
In fact while most of the bright LMXBs in the Milky Way GCs are in metal-rich clusters (e.g. Bellazzini et al. 1995), because these clusters are in the Galactic bulge it has often been argued that the formation of LMXBs in GCs is driven by the galactocentric distance of the clusters (e.g. Grindlay et al. 1987). The possibility that enhanced tidal forces in the inner regions of galaxies promotes the dynamical evolution of clusters, and hence the enhanced formation of LMXBs, provides an attractive physical explanation. In Figure 3 we plot the V-I color distribution (B-I for NGC 1399) of GCs and GC-LMXBs as a function of projected galactocentric distance in our program galaxies. It is clear that GC-LMXBs are located in metal-rich GCs at all galactocentric distances and not just in the inner regions of galaxies. Moreover, there is no significant over-density of GC LMXBs in metal-poor GCs in the inner regions of galaxies, as would be expected if galactocentric distance from the center of the galaxy were a factor. In order to better understand the relative importance of, metallicity, luminosity and galactocentric distance on the LMXB efficiency in GCs we turn to discriminant analysis .

	Discriminant analysis is used to weight and combine the discriminating variables in such a way that the differences between pre-defined groups are maximized (e.g. Antonello \& Raffaelli 1983). Thus each data point is assigned a discriminant score of the form $F = w_1x_1 + w_2x_2 +  ... w_ix_i$
where F is the discriminant score, $w_i$ is the weighting coefficient for variable i, and $x_i$ is the i$^{th}$ discriminating variable, such that the
distribution of discriminant scores of the pre-defined groups is maximally
separated along the axis of this new composite variable. The absolute values of
 the standardized  coefficients,  $w_i$,  reveal the relative
importance of the associated discriminating variables. In certain cases, where the discriminating variables may be correlated,
the absolute value of the structure coefficients - which are the correlations of
each variable with the discriminant function - may give better estimates of the 
significance of each of the variables. 

Using SPSS, we performed discriminant analysis on the LMXB and non-LMXB GC populations with I, V-I (B-I for NGC 1399), and galactocentric distance from the center of the galaxy as the variables. The standardized coefficients and structure coefficients (within brackets) for selected tests are presented in Table 17.   Two random variables, a Gaussian, and a uniform distribution were used to gauge the significance of the results.  We note however that the incompleteness of the faint end of the globular cluster luminosity function has a radial dependence that is both a function of the background light and differences in the exposure times of the WFPC2 images. Restricting the sample to one magnitude past the turnover luminosity of each galaxy, M$_I$$\approx$-8.46 mags (Kundu \& Whitmore 2001) where the completeness is $\sim$100\%, provides a fairer statistical test for galactocentric distance effects.

Discriminant analysis of our sample shows that the luminosity, color and galactocentric distance of GCs can be used to separate the LMXB hosts and non-LMXB GCs in NGC 1399, NGC 3115, NGC 4594 and NGC 4649. The p-value for NGC 3379, which measures the significance of the null hypothesis that there is no discriminating power in the variables based on Wilks' Lambda statistics, is larger than 0.05 suggesting that the discrimination is marginal in this galaxy. This is likely a consequence of the small numbers of GCs and LMXBs in this galaxy. Table 17 reveals that luminosity and color of a GC are  the
most important factors that drive LMXB formation in GCs while galactocentric distance provides negligible discrimination in most cases. 

We note that while the sign of the discriminant weights is inconsequential the {\it relative} signs of the weights for linear discriminant analysis (where there are two predefined groups) indicates the direction of the correlation with the variables. Although galactocentric radius appears to have marginal discriminating power in NGC 4649 (as compared to the random variables), an inspection of the relative signs in concert with the luminosity and metallicity dependence established earlier suggests, surprisingly, that more distant GCs are slightly favored LMXB hosts. This is likely due to contamination of the GC-LMXB sample at larger galactocentric distances by sources associated with the companion spiral galaxy NGC 4647. Of the galaxies in our sample only the discriminant analysis of NGC 3115 reveals evidence of enhanced LMXB formation in GCs closer to the galactic center. However, in this galaxy the discriminants associated with color and galactocentric distance have similar weights which may be due to a higher degree of correlation between these two quantities in this particular galaxy. The GC study of Kundu \& Whitmore (1998) showed that the red, metal-rich GCs in this edge-on S0 galaxy are preferentially located in a thick disk while the blue GCs are associated with a halo. Since GCs at all galactocentric radii in an edge-on disk pass close to the center of a galaxy in projection at some point in their orbit the metal-rich NGC 3115 GCs observed at small projected radii are more likely to be ``contaminated" by the projection of more distant GCs than the more isotropic metal-poor population. However, based on this data set we cannot rule out the possibility that additional dynamical processes due to the presence of a disk in NGC 3115, the only S0 galaxy in our sample, enhances the production of LMXBs in clusters in the inner regions.  In sum our analysis confirms that in galaxies with statistically significant samples of GCs and LMXBs more massive and metal-rich globular clusters are preferred LMXB hosts while there is no convincing evidence that galactocentric radius has any effect on LMXB production in GCs.

\subsection{Implications of the Color Distribution}

These observations make it abundantly clear that red, metal-rich and luminous GCs are more likely to host LMXBs than metal-poor and less massive clusters. A detailed look at the exact correlation between the LMXB hosting probability and cluster metallicity may provide interesting insights about the physics of LMXBs and possibly even the formation history of the host galaxy. For example, the irradiation induced wind model of Maccarone et al. (2004) suggests that the smaller probability of finding GC-LMXBs in metal-poor GCs is due to the shorter lifetime of such binaries. It predicts LMXBs in metal-poor poor GCs, but suggests that there will be much larger fraction in metal-rich clusters at any given time. Ivanova (2006) proposes a different scenario in which the lack of an outer convective zone in metal-poor stars turns off magnetic braking and inhibits mass transfer, predicting a discrete cutoff in the X-ray binary fraction below a fiducial metallicity. While the Ivanova (2006) scenario does not preclude the possibility of LMXBs in metal-poor GCs by other mechanisms such as the formation of ultracompact binaries by direct collisions, if the Ivanova (2006) model is the dominant mechanism for LMXB formation one may expect to see a cutoff in the X-ray binary fraction below a fiducial metallicity.

It is possible that the ratio of 3.4 for the fraction of LMXBs in metal-rich to metal-poor GCs is an underestimate if some of the blue GC-LMXBs are actually in metal-rich but very young ($\lesssim$1 Gyr) GCs. Since LMXB formation may be enhanced in younger stellar populations (e.g. Davies \& Hansen 1998; White \& Ghosh 1998) even a small fraction of such young GCs may increase the apparent LMXB rate in optically blue GCs. However, it would require a fortuitous combination of extremely young ages and less than normal GC mass to place these objects at the observed locations in the color magnitude diagrams (Fig 1). While  studies of the ages and metallicities of GCs in galaxies such as NGC 1399 (Kissler-Patig et al. 1998; Kundu et al. 2005) reveal that there may be a small fraction of intermediate age GCs (3-8 Gyrs) in elliptical galaxies there is no strong evidence of younger GCs in any of our sample galaxies to date. Similarly NGC 4472, which has 7 known LMXBs in blue GCs (KMZ) appears to have only old ($\gtrsim$8 Gyrs) GCs (Hempel et al. 2006; Beasley et al. 2000; Cohen et al. 2003). Future studies  of LMXB formation rates in GCs with known metallicities and ages will help further constrain the exact metallicity dependence.

A curious feature of the NGC 1399 LMXB distribution is that LMXBs appear to be preferentially located in the very reddest clusters of the red GC subpopulation. The median color of red GCs (defined as clusters redder than B-I = 1.96 according to KMM) is B-I = 2.20, while the GCs hosting LMXBs in these clusters have a median B-I color of 2.34, confirming this feature. The existence of such a trend within the red subpopulation implies that the red GCs in NGC 1399 themselves have a large metallicity spread. While the GC systems of most ellipticals are known to be bimodal it has been difficult to establish whether the peaks signify single discrete episode of star formation or mask a range of GCs with different histories due to uncertainties in observational data and conversion from color and absorption line indices to metallicity. Hierarchichal scenarios of galaxy formation predict that giant elliptical galaxies, especially ones located in clusters like NGC 1399, have undergone a series of major and minor mergers accompanied by GC formation (e.g. Beasley et al. 2002). Thus our NGC 1399 data provides an interesting new insight into galaxy formation through the properties of LMXBs. We note however that there is strong evidence of this feature in only this galaxy, which may be due to the larger B-I color baseline of the NGC 1399 data, its rich GC and LMXB systems, or its preferred location at the center of the Fornax cluster. Of the other galaxies in our sample, NGC 4594 shows a slight preference for LMXBs in the reddest GCs of the metal-rich subpopulation (Fig 1). Near-IR observations of a subsample of the NGC 4594 GCs, which provides additional leverage for metallicity constraints, reveals a similar trend (Hempel et al. 2006). On the other hand neither optical (KMZ), nor additional near-IR, observations (Hempel et al. 2006) of NGC 4472 LMXBs shows this effect. It is possible that the galaxy to galaxy variation in the GC-LMXB properties of the red clusters provides an intriguing insight into the differing enrichment histories of the various galaxies. Clearly more observations with larger color baselines that are metallicity sensitive are needed to confirm this preliminary result.

\begin{figure}
\includegraphics[angle=0, scale=0.8]{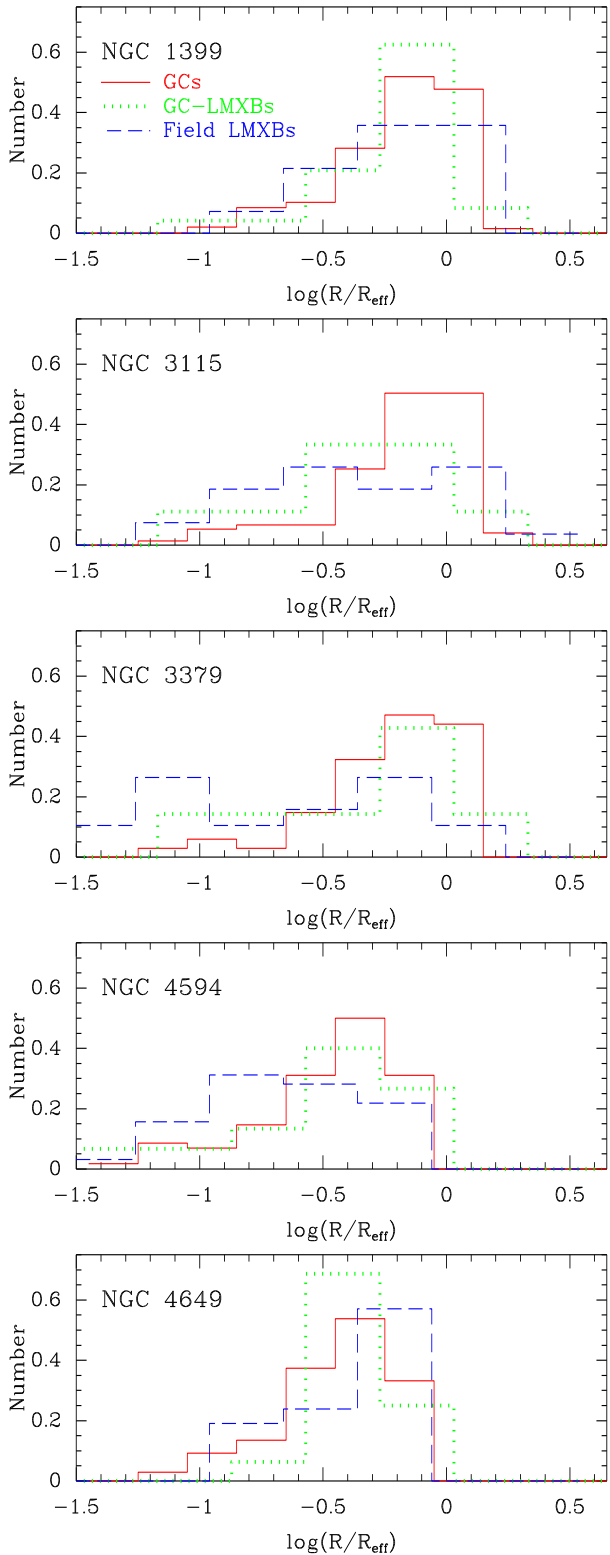}
\caption{The (normalized to 1) radial distribution of GCs, GC-LMXBs, and field LMXBs in our program galaxies.}
\end{figure}

\subsection{Are Field LMXBs an Escaped Cluster Population? }

In many ways the formation of isolated LMXBs in the field is even more uncertain than  LMXBs in GCs. It has long been recognized that it is difficult to keep a low mass component in a binary bound during a supernova explosion due to mass loss from the system and supernova kicks. The possible solutions range from  fine-tuning of the binary and kick parameters (Brandt \& Podsiadlowski 1995; 
Kalogera 1998), common envelope evolution (van den Heuvel 1983; Kalogera \& Webbink 1998) or evolution of an intermediate mass binary system (Podsiadlowski, Rappaport \& Pfahl 2002) that may allow for LMXBs with wider binary separations (Piro \& Bildsten 2002). Another possibility is that most LMXBs are dynamically formed in GCs and subsequently released into the field either due to dynamical ejection (Grindlay \& Hertz 1985; Hut, McMillan, \& Romani 1992) or cluster destruction (Grindlay 1984; Grindlay \& Hertz 1985; Vesperini 2000, 2001; Fall \& Zhang 2001). 

	It is possible to decipher the birthplace of field LMXBs by studying their spatial profiles. The spatial distribution of field sources formed in situ is expected to follow the light (or mass) profile of the host galaxy, while remnants of destroyed globular clusters should be more centrally concentrated due to the higher efficiency of GC destruction in the inner regions of galaxies. Globular cluster systems on the other hand are known to have much larger core radii than that of the underlying light profile (see Ashman \& Zepf 1998 and references therein), and a population of LMXBs ejected from GCs would be expected to follow a similarly more diffuse spatial distribution.

\begin{figure}
\includegraphics[angle=0, scale=0.6]{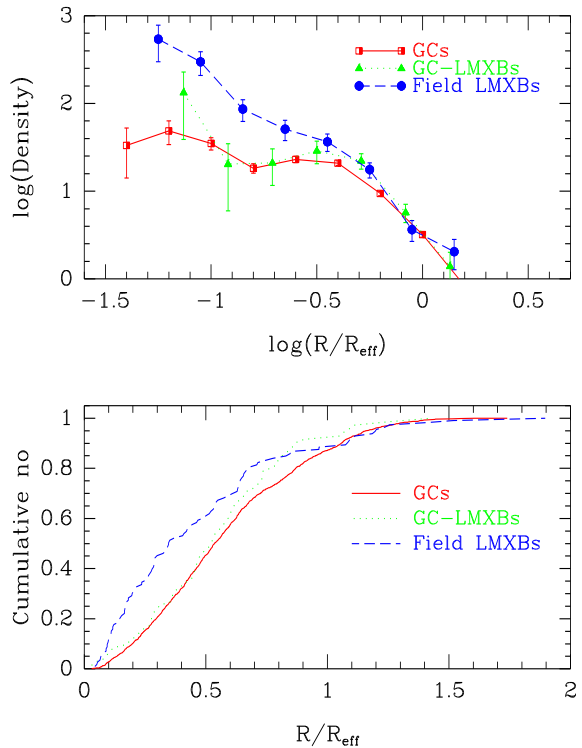}
\caption{Top: The co-added spatial distributions of GCs, GC-LMXBs, and field LMXBs in our galaxy sample. The x-axis has been scaled by the respective de Vaucouleurs et al. (1991) effective radius of each galaxy. The curves have been arbitraily shifted in the y direction such that they all pass through (0,0.5). Bottom: The cumulative radial distributions of GCs, GC-LMXBs, and field LMXBs.  Both plots reveal that field LMXBs are more centrally concentrated than GC-LMXBs. A majority of LMXBs in the field are likely formed in situ and not primarily a dynamically ejected GC population.}
\end{figure}

	Another way to probe the ancestry of field sources is by analyzing the  correlation of LMXB properties with the specific frequency of GCs, an approach used in two recent studies (Irwin 2005; Juett 2005).  Irwin (2005) argues on the basis of the observed trend of specific frequency with integrated LMXB luminosity, and Juett (2005) based on the relationship with the GC-LMXB association rate, that the field sources are likely formed in situ and not ejected from GCs. It is worth noting that both studies probe GC and LMXB properties in a local region in the inner parts of the program galaxies, and implicitly assume that LMXBs ejected from GCs follow the underlying GC profile.  However, the GC spatial distribution only provides an upper bound to the spatial concentration of an ejected LMXB population that is released with small velocities. Any large kick velocities will tend to diffuse ejected LMXBs even further. Moreover, the magnitude of such additional diffusion of an ejected LMXB distribution would depend on the depth and shape of the potential well of the host galaxy. Such a lack of conservation of ejected LMXBs within the field of views of the Juett (2005) and Irwin (2005) studies would mimic the effects of a true field population in their analyses. Moreover as Juett (2005) mentions the supernova kick velocities of LMXBs formed in situ in the field also need to be considered. Large kick velocities of these candidates can also lead to a galaxy potential dependent diffusion of LMXBs to larger galactocentric distances. Thus we attempt to independently test the birthplace of field sources by comparing the spatial profiles.

	The median galactocentric distance of the GC-LMXBs and field LMXBs from the centers of our program galaxies are 51.6'' and 43.6'' in NGC 1399, 34.0 and 27.7 in NGC 3115, 48.4 and 18.6 in NGC 3379, 57.0 and 31.1 in NGC 4594 and 60.5 and 71.6 in NGC 4649, respectively. This suggests that the field population is more centrally concentrated than the GC-LMXBs in all of our program galaxies except NGC 4649. Although we have attempted to eliminate contaminating sources due to the companion spiral NGC 4647 it is likely that X-ray binaries associated with it still contaminate the LMXB list of NGC 4649, especially in the regions furthest from the center of the galaxy, thus accounting for the discrepancy (also see below). In Fig 4 we plot the normalized radial distribution of the GC, GC-LMXB, and field LMXB populations of each galaxy as a function of the effective radius from the center of the galaxy using the RC3 effective radii (de Vaucouleurs et al. 1991). In each galaxy, there is evidence that the field LMXBs are more centrally concentrated than the GC-LMXBs. Even in NGC 4649 field LMXBs outnumber GC-LMXBs in the innermost part of the galaxy, providing further evidence that the apparently anomalous nature of the median radial galactocentric distances is due to residual contamination by X-ray sources associated with NGC 4647 in the outer regions of our images. The GC distribution plots the clusters brighter than the turnover magnitude, and is effectively complete at all galactocentric radii.  The LMXB detection rate may be affected by incompleteness, especially in the innermost regions. However since both GC-LMXBs and field LMXBs are affected similarly by X-ray incompleteness effects, the relative difference between the profiles is a robust measure. Thus the increase in the relative density of field vs. GC LMXBs at small galactocentric radii indicates that LMXB ejection from GCs is not the primary source of field LMXBs.

In the upper panel of Fig 5 we plot the co-added radial spatial distributions of GCs, GC-LMXBs, and field LMXBs, arbitrarily normalized in the vertical direction so they all pass through log (R/R$_{eff}$) = 0. The lower panel traces the cumulative spatial profiles of these populations. Fig 5 confirms that field LMXBs are more centrally concentrated than GC LMXBs. 

Previous studies of the radial profiles of LMXBs have provided mixed results. Our analysis of NGC 4472 (KMZ, MKZ) showed no statistically significant evidence of differences in the radial profiles of field and GC-LMXBs, which we interpreted as evidence of a population of ejected GC sources. Although the NGC 4472 observations covered a larger radial range than this data set the outer regions with low LMXB densities likely contributed significant noise to the profile of this single galaxy. Restricting the NGC 4472 observations to objects closer than 70'' yields a median galactocentric distance of 41.2'' for field LMXBs and 50.4'' for GC-LMXBs, consistent with the results of this study. The trend is unaffected by the exact choice of cutoff radius. Kim et al (2006) on the other hand suggested that {\it both} field-LMXBs and GC-LMXBs are more centrally concentrated than the GC system in their program galaxy. They attributed the steeper profile of the GC-LMXBs to enhanced dynamical LMXB production in the inner regions of galaxies. As discussed above we find no convincing evidence for such an effect either in Fig 3 or in our discriminant analysis. Kim et al. (2006) analyze a combination of HST and ground-based data in their analysis. In addition to the possible effects of low LMXB density in the outer regions of their data sets, the likely larger contamination rate of their ground-based optical sample may affect their GC profile. Similarly it is possible that our LMXB observations may be affected by incompleteness in the innermost regions. We note however that the Kim et al. (2006) data do show that the LMXBs in metal-poor GCs are distributed more diffusely than the ones in metal-rich GCs.

Thus we conclude on the basis of our observations  that field LMXBs are more centrally concentrated than GC-LMXBs and are likely associated with the underlying diffuse stellar component. This is consistent with the conclusions of Irwin (2005) and Juett (2005).
 However, we note that none of the three methods can eliminate the possibility that the field sources are remnants of destroyed GCs. Since the modification of the spatial distribution of GCs occurs throughout the lifetime of a galaxy, although  particularly efficiently early in its history, such a destroyed GC remnant population would either have to survive over a Hubble time due to a low duty cycle, or have a long binary evolution phase before turning on. It is unlikely that GCs that have undergone accelerated dynamical evolution would preferentially produce such a population of binaries with the generally larger separations required in such a scenario. 
Thus we conclude that the preponderance of evidence points towards in-situ formation for a majority of the field LMXBs.

\subsection{ Global Fractions and Variations with Host Galaxy}

The fraction of LMXBs associated with GCs in our program galaxies is 63\% in NGC 1399, 25\% in NGC 3115, 27\% in NGC 3379, 32\% in NGC 4594 and 46\% in NGC 4649, confirming the galaxy to galaxy variation noted in previous studies (e.g. MKZ; Kim et al. 2006). However, the range is somewhat smaller than the 20\%-70\% quoted in  many papers. The upper end of this figure comes from the Angelini et al. (2001) analysis of NGC 1399. We note that Angelini et al. (2001) have made no color selection for GCs on the optical lists used to match X-ray and optical sources. This likely accounts for the discrepancy as we show above that a handful of sources in NGC 1399 are likely to be associated with non-GC contaminants. The lower bound comes from the NGC 4697 analysis of Sarazin et al. (2000, 2001). The shallow, single filter image used by Sarazin et al. (2000) identifies only a fraction of the GCs in the field of view. A more complete statistical analysis by Sarazin et al. (2003) that accounted for selection effects suggested that about 40\% of the LMXBs in this galaxy are associated with GCs. We note that the $\approx$10\% figure for GC-LMXB matches for the Galaxy (e.g. Liu et al. 2001) is the global fraction. Restricting the Milky Way data to the typical several kpc field of the HST data of other galaxies (or limiting the data to a similar effective radius) would yield a considerably higher GC-LMXB fraction because most of the GC-LMXBs (and the metal-rich GCs) are in the inner regions of the Galaxy.

\begin{figure}
\includegraphics[angle=0, scale=0.8]{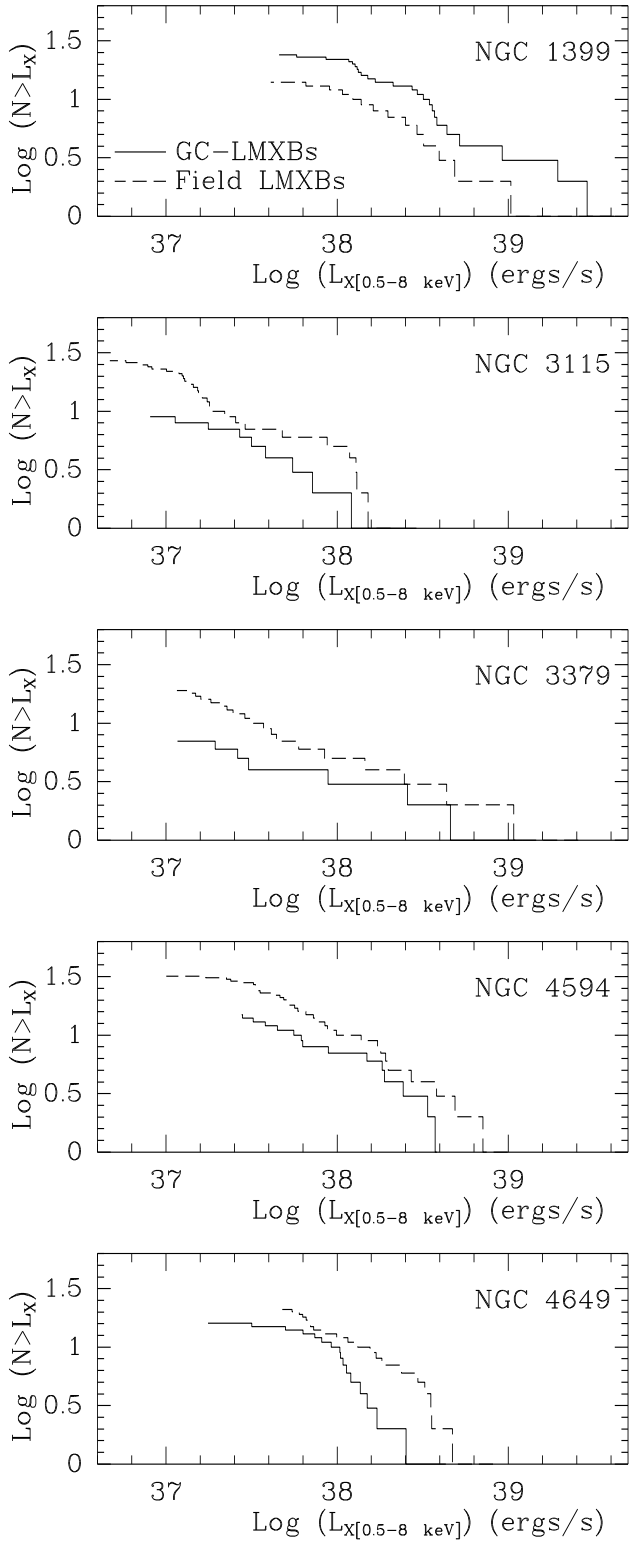}
\caption{ The luminosity functions of LMXBs associated with globular clusters and field LMXBs for sources within the HST-WFPC2 field of view.}
\end{figure}

Previous studies  have noted that the fraction of LMXBs in GCs appear to be a function of galaxy type (MKZ; Sarazin et al. 2003; Irwin 2005) with early type galaxies (NGC 1399, NGC 4472) having a larger fraction of LMXBs in GCs than S0 galaxies (NGC 3115, NGC 1553), which in turn have greater GC-LMXB fractions than disk galaxies (Milky Way, M31). The relatively low fraction of LMXBs associated with GCs in the nearby elliptical galaxy NGC 3379 does not appear to support this trend. However, NGC 3379 does have a relatively low GC specific frequency for an elliptical galaxy and hence agrees with the general characteristic of increasing GC-LMXB association rate with specific frequency found in previous studies.

	The global fractions of LMXB-GC associations can also be modulated by variations in the fraction of GCs hosting LMXBs in different galaxies. Previous studies suggest that this number is a relatively constant 4\% for LMXBs brighter than $\sim$10$^{37}$ ergs s$^{-1}$. For LMXBs brighter than 4$\times$10$^{37}$ ergs s$^{-1}$ in our samples the fraction of GCs hosting LMXBs LMXB is 4\% in NGC 1399, 3\% in NGC 3115, 5\% in NGC 3379, 6\% in NGC 4594 and 4\% in NGC 4649.
The efficiency if of course higher when pushed to lower luminosities, especially for the galaxies that are nearby, but the $\sim$10$^{37}$ ergs s$^{-1}$ cutoff represents a luminosity at which all the galaxies our sample are generally complete. Given the preference of LMXBs for metal-rich clusters, and the tendency of metal-rich GCs to be more centrally concentrated than the metal-poor ones these numbers are also likely to be smaller for observations that probe larger spatial scales. 

	Any conclusions based on the matching fractions of LMXBs should be made with caution because of selection biases. In Fig 6 we plot the cumulative luminosity functions of field and GC-LMXBs within the HST field of view in our program galaxies. The significantly different threshold luminosities, which are a function of distance and observing time, are apparent. While the field and GC LMXB luminosity functions seem to be broadly similar it is not obvious that they are identical in each galaxy. For example if the observations of the nearby S0 galaxy NGC 3115 were limited to $\approx$2.5$\times$10$^{37}$ ergs s$^{-1}$, similar to the limit for the more distant ellipticals, the fraction of LMXBs in GCs would nearly double from the 25$\%$ figure quoted above, affecting some of the conclusions of studies based on matching fractions such as Juett (2005) and Irwin (2005).

 The behavior of the X-ray luminosity functions at the faint end may provide interesting clues to the formation of LMXBs in the field. If the formation channel of field LMXBs (e.g. intermediate mass binary evolution, common envelope evolution, or preferred kicks) is different from the dynamical formation of LMXBs in clusters, the orbital parameters, recurrence timescales, and luminosity functions of the two populations is unlikely to be identical. Moreover, observations of the Sculptor Dwarf Spheroidal galaxy (Maccarone et al. 2005) suggest that the duty cycle of field and GC-LMXBs might be quite different, implying a difference in the underlying binary properties. There is evidence that differences in binary orbital period distributions lead to differences in in duty cycles and peak luminosities, and hence luminosity functions (e.g. Chen, Shrader, \& Livio 1997; Portegeis Zwart, Dewi, \& Maccarone 2005). While the X-ray luminosity functions at the faint end of Fig 6 hints at  differences between the field and GC-LMXBs, the small number statistics and limited depth of these observations do not allow for any firm conclusion. Deep observations of the X-ray luminosity function of nearby galaxies (e.g. Kim et al. 2006a) will help resolve this issue.

	There are conflicting claims in the literature about the relative distribution of luminous LMXBs in the field and in GCs. While Angelini et al. (2001) show that LMXBs in GCs are on average brighter than the field LMXBs, Sarazin et al. (2003) suggest that the brightest LMXBs tend to avoid GCs. Kim et al. (2006) on the other hand find roughly equal numbers of bright LMXBs in GCs and in the field and argue that the X-ray luminosity functions in GCs and the field are consistent. Our data shows that the GC-LMXBs are on average brighter than the field ones in NGC 1399 (Fig 6) but are fainter than field LMXBs in all the other galaxies. However, we note that NGC 1399 is the only galaxy in our sample in which more than half the LMXBs within the optical field are in GCs. Thus our observations, and the previously published ones, are consistent with the explanation that the GC and field LMXBs have similar LFs and the differences  in the mean luminosities of the two populations in various galaxies is largely  reflective of the relative sample sizes drawn from a power law luminosity function. This broadly consistent explanation is somewhat complicated by the KMZ observation that the mean L$_X$ of the GC sources in NGC 4472 is marginally brighter than the mean luminosity of the field LMXBs even though only $\approx$40\% of the LMXBs are associated with GCs. However, the median luminosity provides a better test of possible differences in the luminosity distributions of X-ray binaries. The typical cumulative X-ray luminosity function L(N$>$$L_X$) is well fit by an expression of the form:

\begin{equation}
Log(N>L_X) = Log(N>L_{X_{0}}) + \alpha [ Log(L_X) - Log(L_{X_{0}}) ]
\end{equation}

	If L(N$>$$L_{X_0}$) is the faintest source in the entire sample and L$_{X_{0}}$ is its corresponding luminosity then the median L$_X$ of the distribution, and any subsample of the distribution with the same low luminosity cutoff, is L$_{X_{0}}$ - Log(2)/$\alpha$. Therefore if we assume that field and GC LMXBs distributions follow the same power law exponent $\alpha$, and reasonably assume that they have the same completeness limit, then the median L$_X$'s of the two populations should be equal, irrespective of the exact value of $\alpha$. The median logarithmic luminosity of the GC and field sources in our sample galaxies are 38.46:38.30 in NGC 1399, 37.5:37.19 in NGC 3115, 37.52:37.51 in NGC 3379, 37.80:37.77 in NGC 4594, and 38.02:38.07 in NGC 4649. Thus except for NGC 4649, where the LMXB population is likely contaminated by X-ray sources associated with its neighboring disk galaxy companion, there is mildly suggestive evidence that GC-LMXBs are slightly brighter than their field counterparts. As a measure of the uncertainty in the median luminosities of the field and LMXB GCs we note that there are 4 sources in the combined GC and field LMXB sample of NGC 1399, 8 in NGC 3115, 0 in NGC 3379, 3 in NGC 4594 and 6 in NGC 4649 that lie between the median GC-LMXB and field LMXB luminosities of the respective galaxies. We note that since the field LMXBs are concentrated towards the centers of the respective galaxies the preferential incompleteness due to hot gas in the central regions of some of the galaxies likely causes an overestimate of the median luminosity of the field sources, and 
hence suppresses the magnitude of the effect. It is possible that the small difference in LFs is due to different LMXB formation mechanisms at work in the two locales. For example, it is possible that dynamically formed GC-LMXBs are ultracompact X-ray binaries (e.g. Bildsten \& Deloye 2004) while field LMXBs may be wide binaries with red giant companions (e.g. Piro \& Bildsten 2002). We explore another possible reason for this effect in the next section.

\begin{figure}
\includegraphics[angle=0, scale=0.5]{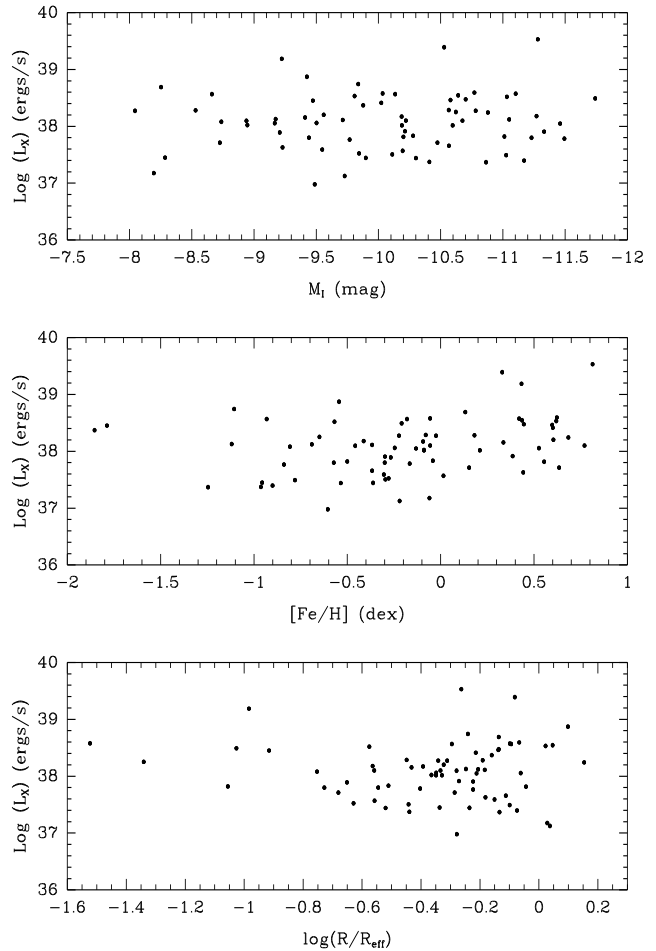}
\caption{ Top: The luminosity of GC-LMXBs in our sample as a function of cluster absolute magnitude. The Tonry et al. (2001) distances to our program galaxies have been used to convert apparent magnitudes to absolute magnitudes. Middle: The variation of GC-LMXB luminosity as a function of cluster metallicity. The broadband colors of the GCs have been converted to metallicity using the Smits et al (2006) color-metallicity relations. Bottom: The luminosity of the GC-LMXBs vs. their galactocentric distance. The galactocentric distance of each GC-LMXB has been normalized by the de Vaucouleurs et al. (1991) effective radius of its host galaxy. While there is no obvious strong correlation between the luminosity of the LMXBs and any of the porperties of their host GCs, the very brightest X-ray sources appear to prefer high-mass and metal-rich globular clusters. }
\end{figure}

\subsection{Correlations with LMXB Luminosity and the Implications on Black Hole LMXBs}

	In Fig 7 we plot the luminosity of the LMXB sources in GCs as a function of the absolute I magnitude of the candidates, the metallicity [Fe/H], and the distance from the centers of the respective galaxies. We have used the linear (B-I) and (V-I) color vs. [Fe/H] relations derived by Smits et al. (2006) to convert the colors to [Fe/H], and Tonry et al (2001) distances to convert the I band magnitudes of the GCs to absolute magnitudes. There is no obvious correlation between the L$_X$ and any of the properties of the host GCs in Fig 7. However, the brightest LMXBs in our sample appear to be preferentially located in metal-rich and luminous GCs. The most luminous LMXBs are particularly interesting because they are well over the L$_X$$\approx$3$\times$10$^{38}$ ergs s$^{-1}$ Eddington luminosity for spherical accretion on to a 1.4 M$_\odot$ neutron star, and hence may be black hole accretors.   In light of the possible preferred GC hosts of the brightest LMXBs, and the strong dependence of LMXB hosting frequency on GC mass and metallicity, it is worth considering in detail whether it is viable that L$_X$$>$L$_{Edd(NS)}$ GC-LMXBs are multiple neutron stars rather than black holes.

\begin{figure}
\includegraphics[angle=0, scale=0.5]{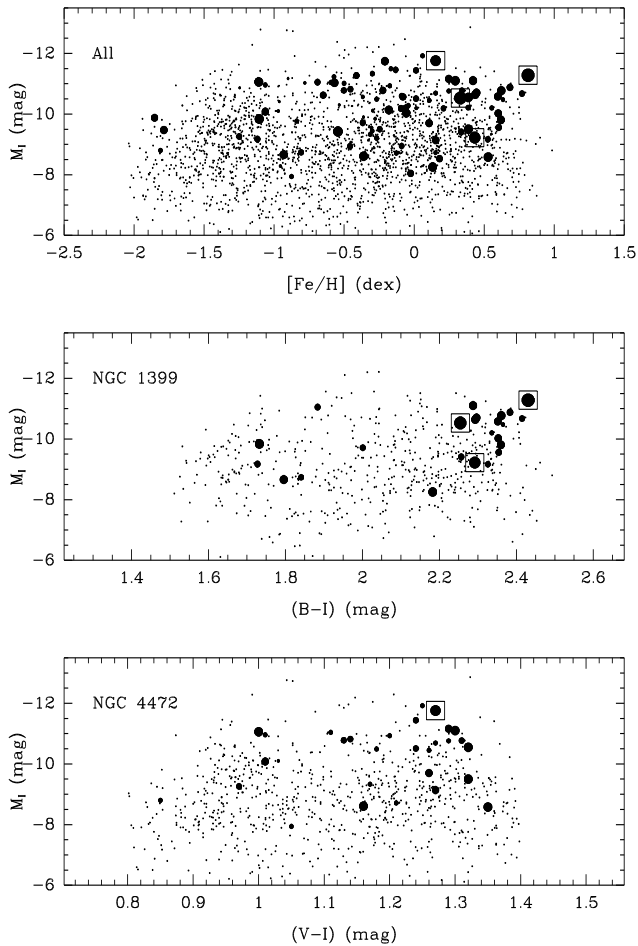}
\caption{Top: The metallicity-absolute magnitude distribution of the GCs in our five program galaxies and NGC 4472 (from Maccarone, Kundu \& Zepf 2003). The filled circles represent LMXB-GCs. The size of the points is proportional to the logarithm of the LMXB luminosity. Boxes mark the sources brighter than L$_X$=10$^{39}$ ergs s$^{-1}$. Middle: The color-magnitude distribution of GCs and GC-LMXBs in NGC 1399, with the size of the filled dots proportional to the logarithm of the luminosity. Bottom: The NGC 4472 color-magnitude and X-ray luminosity plot. The brightest LMXBs appear to prefer metal-rich, high mass clusters. There is a non-negligible chance that at least some of the brightest GC-LMXBs are superpositions of multiple LMXBs.}
\end{figure}

	In the top panel of Fig 8 we plot the absolute magnitude vs. metallicity distributions of all the galaxies in our sample, along with the data for NGC 4472 from MKZ. The filled dots represent LMXB hosts, with the size of the points proportional to the logarithm of the X-ray luminosity. The 4 LMXBs brighter than L$_X$=3$\times$10$^{39}$ ergs s$^{-1}$ in our sample (three in NGC 1399 and one in NGC 4472) are marked by boxes. These sources appear to be found preferentially in metal-rich GCs, and three of the four reside in some of the most massive GCs hosting LMXBs. The color magnitude distributions of the GCs and GC-LMXBs and their relative X-ray luminosities in NGC 1399 and NGC 4472 are isolated in the lower panels of Fig 8. The brightest source in NGC 4472 is in one of the most massive GCs in the sample, while one of the probable black holes in NGC 1399 is in one of the most metal-rich GCs. As we have established that LMXBs are preferentially found in high mass, metal-rich GCs it seems plausible that some of the brightest LMXBs may be superpositions of multiple sources. The fraction of LMXBs residing in a globular cluster system can be written as:

\begin{equation}
\frac{N_{GC-LMXB}}{N_{GC}} = \frac{ {\int}{\int} N(Z,M) \times f(Z,M) dZ dM} {\int \int N(Z,M) dZ dM}
\end{equation}

	where Z is the metallicity of a cluster, M is the GC mass, N(Z,M) is a function that describes the number distribution of GCs as a function of Z and M, and f(Z,M) is a function that describes the probability that a cluster with metallicity Z and mass M hosts a LMXB. Using the data presented in this paper, and the NGC 4472 observations from MKZ, Smits et al. (2006) showed that the probability that a GC hosts a LMXB can be approximated by:

\begin{equation}
f(Z,M) \propto Z^{(0.25\pm0.03)} \times M^{(1.03\pm0.12)}
\end{equation}

	While it is possible to probe the functional form of f(Z,M) using the combined sample, the different X-ray detection limits in the various galaxies require that each host be investigated separately to determine the constant of proportionality in the expression above. Since NGC 1399 and NGC 4472 are the only two galaxies with GC-LMXBs brighter than  L$_X$=3$\times$10$^{39}$ ergs s$^{-1}$ we study these. Using the color-metallicity relation for GCs from Smits et al. (2006) and assuming that the mass-to-light ratio of GCs in our sample is constant we can rewrite eqn. 3 in terms of the observables.

\begin{eqnarray}
f(Z,M) \propto 10^{0.69(B-I)} \times 10^{-0.41 m_I} \\
f(Z,M) \propto 10^{1.17(V-I)} \times 10^{-0.41 m_I}
\end{eqnarray}

For NGC 1399 we solve the expression:

\begin{equation}
\frac{N_{GC-LMXB}}{N_{GC}} = \frac{24}{548} =  \sum C \times 10^{0.69(B-I)} \times 10^{-0.41 m_I} 
\end{equation}

The constant C is found to be 2.99$\times$10$^{10}$. For the 37 GCs with (B-I)$>$2.2 and m$_I$$<$-10 this predicts $\approx$6 GC-LMXBs within this set of objects. The probability that there is at least one GC with multiple LMXBs is $1-\frac{36!}{31!\times37^5}$, or 30\% chance. But there are actually 12 observed GC-LMXBs in these 37 luminous, metal-rich clusters, of which 10 are brighter than L$_X$=10$^{38}$ ergs s$^{-1}$. Thus there is a $1-\frac{36!}{25!\times37^{11}}$, or 87\% chance that there is more than one LMXB in at least one GC and $1-\frac{36!}{27!\times37^9}$, or 74\% chance that at least one GC has more than one L$_X$$>$10$^{38}$ ergs s$^{-1}$ source. There is a 24\% chance that there are three LMXBs in one GC. While multiple LMXBs and qLMXBs have been discovered in a number of Galactic GCs (e.g. White \& Angelini 2001; Charles, Clarkson, \& van Zyl 2002; Heinke et al. 2003), these are all much fainter than the candidates being considered here. We note that our computation here is specifically for sources brighter than L$_X$=10$^{38}$ ergs s$^{-1}$, in a sample of LMXBs with a detection limit of  L$_X$=2.5$\times$10$^{37}$ ergs s$^{-1}$.

The probability of superposed sources derived here is a lower limit because if there are indeed some incidences of multiple sources then the number of discrete LMXBs is being under-counted. The analytically estimated number of LMXBs is likely too low in part because of the apparently stronger than normal metallicity dependence of LMXB hosts in NGC 1399 (\S3.2), and non-linearities in the color-metallicity relation. Since the assumption of a power law dependence of the LMXB hosting probability of a globular cluster is arbitrary this may also indicate that the metallicity correlation in all galaxies actually has a much steeper dependence (e.g. an exponential). Moreover, there is likely to be an additional correlation with GC sizes since the stellar interaction rate, and hence dynamical LMXB formation rate, is expected to correlate with GC density. KMZ showed that GCs with smaller half-light radii are slightly favored LMXB hosts. However, it is the core radius that is most important to the formation of LMXBs in GCs. The core radii of GCs at the distances of these galaxies is only a small fraction of a pixel and is extremely challenging to measure. While Jordan et al. (2004) claim to have measured the core radii of M87 GCs and derived the interaction rate, we suggest in Smits et al. (2006) that these measurements are consistent with a random distribution of sizes, as would be the case if there were no constraint on the core radii.

To investigate the probability of LMXB superpositions in NGC 4472 we solve:
\begin{equation}
\frac{N_{GC-LMXB}}{N_{GC}} = \frac{30}{825} =  \sum C \times 10^{1.17(V-I)} \times 10^{-0.41 m_I} 
\end{equation}

	This yields a value of 2.10$\times$10$^{10}$ for the constant C and predicts $\approx$10 LMXBs in the 55 GCs with (V-I)$>$1.2 and m$_I$$<$-10. Thus there is a 60\% chance that there are multiple LMXBs in at least on GC. There are actually 11 bright LMXBs in these clusters giving a comparable probability for multiple sources. This also suggests that the greater numbers of bright LMXBs in NGC 1399 than predicted by equation 6 are likely due to the unusual metallicity effects in this galaxy. 

	Given the fact that we have ignored any correlation with core radius, and the exponents on both variables in f(Z,M) is likely underestimated because it assumes that each GC-LMXB in our sample hosts a single X-ray binary, the probabilities derived above should be considered lower limits for superpositions in high mass, metal-rich GCs. Thus it seems likely that at least a part of the reason for the slightly higher median luminosities of GC-LMXBs as compared to the field sources is due to multiple bright LMXBs in some of the clusters.  On the other hand, assuming a 3:1 ratio of LMXBs in metal-rich vs. metal-poor GCs yields only a $\sim$3\%-15\% chance of multiple LMXBs in the most massive, metal-poor GCs in NGC 1399 and $\sim$10\% in NGC 4472. Moreover, measurements of the half-light radii of extragalactic GCs reveal that metal-poor GCs are on average larger than metal-rich GCs (Kundu \& Whitmore 1998) thereby possibly increasing the contrast in the LMXB superposition probabilities in metal-rich and metal-poor GCs (if this is indicative of the core interaction rate as weakly hinted by Kundu et al. 2000). These observations lead us to the conclusion that luminous LMXBs in metal-poor GCs are more likely to be black hole candidates while the ones in metal-rich GCs, such as the ones studied by Irwin (2006), may appear to be stable because of the non-negligible probability of superposition of multiple LMXBs in some of the GCs. Therefore, the presence of black holes in extragalactic GCs can be confirmed only if there are large temporal luminosity variations consistent with a single source (Kalogera, King, \& Rasio 2004).
We have recently discovered just such a black hole in a metal-poor globular cluster in NGC 4472 (Maccarone et al. 2007). We note that while our analysis suggests that the bright LMXBs in metal-rich GCs are superpositions it does not imply that these are superpositions of neutron star LMXBs. These GCs may host multiple black holes. Black holes in more metal-rich systems are expected to have lower masses due to the metallicity dependent effect of stellar mass loss (e.g. Fryer \& Kalogera 2001). Such lower mass black holes might not decouple from the stars due to the Spitzer instability that is invoked to explain the ejection of most black holes from globular clusters (Kulkarni, Hut \& McMillan 1993; Sigurdsson \& Hernquist 1993). Thus metal-rich GCs may retain black holes more effciently than metal-poor ones (Vicky Kalogera, private communication).

	While we have concentrated on the brightest LMXB sources in this dicussion we note that due to the power law nature of the LMXB luminosity function there are likely to be a larger number of superpositions of fainter LMXB sources in globular cluster systems with high metallicity clusters. Thus, the apparent similarity of the luminosity function of LMXBs in the field and GCs in combined samples (e.g. Kim et al. 2006) may conceal underlying metallicity dependent variations in the GC-LMXB luminosity function. For example, our suggestion of multiplicity of LMXBs in metal-rich GCs predicts that the most metal-rich cluster systems should have inordinately bright total X-ray luminosities because the effects of superpositions are folded in. This would lead to a bright tail in the GC-LMXB X-ray luminosity function. The NGC 1399 system shows just such a tail in Fig 6. Thus we suggest that the shape of the bright end of the metallicity distribution of GC-LMXBs may be indicative of the metallicity distribution of the underlying GCs, with the most metal-rich systems having the largest tails. If the metal-rich GCs indeed host multiple LMXBs we also expect that such GC-LMXBs will on average show less variability than more metal-poor sources.

\section {Conclusions}

	We have analyzed 1356 globular clusters in HST-WFPC2 images of 5 elliptical and S0 galaxies, and 641 LMXBs in Chandra X-ray images of these candidates. Of the 186 LMXBs within the WFPC2 fields, 68 are in globular clusters. Figs 1 and 2 show that LMXBs are preferentially associated with bright, metal-rich clusters in these galaxies with known bimodal GC metallicity distributions, confirming the trends seen in previous studies. Metal-rich clusters are 3.4 times as likely to host LMXBs as blue metal-poor ones. This is similar to the ratio in our previous analysis of NGC 4472 (KMZ) but higher than that measured in other surveys (Sarazin et al. 2003; Kim et al. 2006), likely because of larger contamination and/or the choice of galaxies with intermediate age clusters in the latter studies. 

The LMXBs in NGC 1399 reveal a strong preference for the reddest, most metal-rich clusters, suggesting that there is a significant metallicity spread in the red subpopulation indicative of multiple star formation episodes within the red peak. The range of this effect seen in this data set and supporting near-infrared observations ranges from the strong evidence seen in NGC 1399, through NGC 4594 which also might show a similar correlation, to NGC 4649 in which it is not clear, and NGC 4472 in which it does not appear. It is not clear whether this is due to different enrichment (and hence formation) histories among the galaxies or possible variations due to small numbers or other characteristics of the individual data sets. 

We find no statistically convincing evidence that there is any correlation between the galactocentric distance of a GC and its probability of hosting a LMXB either in Fig 3 or by discriminant analysis.  Figs 4 and 5 reveal that it is unlikely that the field sources are predominantly made in GCs and injected into the field either by dynamical ejection, or by cluster destruction. Thus we conclude that field sources are associated with the diffuse component of the galaxy and likely formed in situ. This independently confirms the conclusions of Irwin et al (2005) and Juett et al (2005) based on specific frequency arguments. 
While there is no strong correlation between the luminosity of GC-LMXBs and the metallicity, mass, or galactocentric distance of the host GCs there are intriguing hints in Figs 7 and 8 that the brightest GC-LMXBs are in the most luminous, metal-rich globular clusters. We show in \S3.5 that there is a reasonable probability that some of the brightest GCs may harbor multiple bright LMXBs. There is much higher probability that some of the luminous black hole LMXB candidates in metal-rich GCs are superpositions of multiple LMXBs while the corresponding luminous LMXBs in metal-poor GCs are more likely to be bona fide black hole candidates. The only convincing way to prove the existence of a black hole LMXB in a globular cluster is to detect large amplitude variations that rules out the possibility of multiple bright neutron star LMXBs.  If the interpretation of multiple LMXBs in metal-rich clusters is correct our study implies that the shape of the X-ray luminosity function of GC-LMXBs at the brightest end hints at the peak metallicity of a GC system, with most metal-rich GC systems having a large tail. In this scenario metal-rich GCs should also show less X-ray variability than metal-poor ones.

\acknowledgements
SEZ and AK gratefully acknowledge support from NASA-LTSA grants NAG5-11319 and NAG5-12975 and Chandra grant AR-6013X. We thank the referees for the comments that helped improve this paper. We also thank Vicky Kalogera for several helpful discussions.

\phantom{a}
\clearpage
\begin{deluxetable}{cccccccc}
\tablenum{1}
\tablecaption{Galaxy Sample }
\startdata
\tableline
 	& 	& \multicolumn{2}{c}{\underline{Chandra Data}} & \multicolumn{4}{c}{\underline{HST Data}}\\
Galaxy	& Distance (m-M)\tablenotemark{1} & Exposure time & Obs. date	& B	& V 	& I  & Obs. date\\
	& (mags)		& (ksecs) &  	& (secs)	& (secs)	& (secs) &  \\
\tableline
\tableline
\\
NGC 1399 & 31.50 	& 56.7	& Jan 2000	& 5200		& \nodata	& 1800 &	Jun 1996\\
NGC 3115 & 29.93	& 37.4	& Jun 2002	& \nodata	& 1050		& 1050 &	Nov 1994\\
NGC 3379 & 30.12	& 31.9	& Feb 2001	& \nodata	& 1500		& 1200 &	Nov 1994\\
NGC 4594 & 29.95	& 18.8	& May 2001	& \nodata	& 1200		& 1050 &	Dec 1994\\
NGC 4649 & 31.13	& 47.4	& Apr 2000	& \nodata	& 2100		& 2500 &	Apr 1996\\											\enddata
\tablenotetext{1} {SBF distances from Tonry et al. 2001} 
\end{deluxetable}

\begin{deluxetable}{llllll}
\tablewidth{0pt}
\tablenum{2}
\tablecaption{NGC 1399 X-ray Binaries}
\startdata
    \\
\tableline
\tableline
	  No.& Chandra ID		    &	RA(J2000)  & Dec(J2000)  & L$_X$ (ergs s$^{-1})$   & HST-FOV  \\
\tableline
           1\tablenotemark{1} &       CXOKMZ J033829.0-352701 &   3:38:29.03 &  -35:27:01.2 & 7.18E40 & Y \\
           2 &       CXOKMZ J033828.6-352708 &   3:38:28.67 &  -35:27:08.9 & 1.54E39 & Y \\
           3 &       CXOKMZ J033830.0-352655 &   3:38:30.02 &  -35:26:55.4 & 3.00E38 & Y \\
           4 &       CXOKMZ J033828.2-352711 &   3:38:28.22 &  -35:27:11.5 & 1.20E38 & Y \\
\tableline
\enddata
\tablecomments{The complete version of this table is available in the electronic version of the Journal}
\end{deluxetable}

\begin{deluxetable}{rrrrrrr}
\tablewidth{0pt}
\tablenum{3}
\tablecaption{NGC 1399 Globular Clusters}
\startdata
\\
\tableline
\tableline
	  No.& RA(J2000)	    & Dec(J2000)  & B (mag)  & I (mag)  & B-I (mag)  \\
\tableline
  1\tablenotemark{1} & 3:38:29.23 & -35:27:00.9 & 23.57$\pm$0.2 & 21.51$\pm$0.3 & 2.06$\pm$0.36 \\
  2\tablenotemark{1} & 3:38:28.90 & -35:27:04.6 & 23.57$\pm$0.14 & 21.95$\pm$0.3 & 1.62$\pm$0.33 \\
  3\tablenotemark{1} & 3:38:28.63 & -35:26:57.6 & 23.51$\pm$0.08 & 21.57$\pm$0.12 & 1.94$\pm$0.14 \\
  4\tablenotemark{1} & 3:38:29.37 & -35:27:06.4 & 24.34$\pm$0.12 & 22.42$\pm$0.17 & 1.92$\pm$0.21 \\
\tableline
\enddata
\tablecomments{The complete version of this table is available in the electronic version of the Journal}
\end{deluxetable}

\begin{deluxetable}{lllllll}
\tablewidth{0pt}
\tablenum{4}
\tablecaption{NGC 1399 GC-LMXBs}
\startdata
    \\
\tableline
\tableline
 Chandra ID		    &	XID  & OID  & L$_X$ (ergs s$^{-1})$   & B (mags) & I (mags) & B-I (mags)  \\
\tableline
CXOKMZ J033828.6-352708  &  2  &   7  &  1.54E39  & 24.57$\pm$0.1 & 22.28$\pm$0.11 & 2.29$\pm$0.15 \\
CXOKMZ J033828.2-352711  &  4  &  30  &  1.20E38  &  24.6$\pm$0.07 & 22.76$\pm$0.09 & 1.84$\pm$0.11 \\
CXOKMZ J033829.2-352630  &  8  &  88  &  1.43E38  & 24.35$\pm$0.05 & 22.09$\pm$0.04 & 2.26$\pm$0.07 \\
CXOKMZ J033831.7-352644  & 12  & 124  &  1.26E38  & 23.24$\pm$0.02 & 20.82$\pm$0.01 & 2.41$\pm$0.02 \\
CXOKMZ J033832.1-352705  & 13  & 130  &  1.59E38  & 24.29$\pm$0.03 & 21.94$\pm$0.03 & 2.35$\pm$0.05 \\
CXOKMZ J033832.3-352701  & 15  & 146  &  3.68E38  & 24.63$\pm$0.05 & 22.84$\pm$0.06 &  1.8$\pm$0.07 \\
CXOKMZ J033826.4-352634  & 16  & 153  &  5.15E37  & 23.39$\pm$0.02 & 21.03$\pm$0.01 & 2.37$\pm$0.02 \\
CXOKMZ J033832.6-352705  & 23  & 170  &  3.39E39  & 22.65$\pm$0.01 & 20.22$\pm$0.01 & 2.43$\pm$0.01 \\
CXOKMZ J033832.6-352652  & 25  & 177  &  1.34E38  & 24.05$\pm$0.03 & 22.33$\pm$0.04 & 1.73$\pm$0.05 \\
CXOKMZ J033832.8-352658  & 27  & 184  &  5.53E38  & 23.39$\pm$0.02 & 21.66$\pm$0.02 & 1.73$\pm$0.03 \\
CXOKMZ J033832.3-352729  & 30  & 204  &  2.58E38  & 23.83$\pm$0.03 & 21.48$\pm$0.02 & 2.35$\pm$0.03 \\
CXOKMZ J033833.0-352651  & 32  & 209  &  1.32E38  & 22.33$\pm$0.01 & 20.45$\pm$0.01 & 1.88$\pm$0.01 \\
CXOKMZ J033832.4-352734  & 35  & 228  &  1.29E38  & 23.79$\pm$0.02 & 21.79$\pm$0.02 &   2.$\pm$0.03 \\
CXOKMZ J033833.7-352658  & 41  & 263  &  2.90E38  & 23.27$\pm$0.02 & 20.92$\pm$0.01 & 2.35$\pm$0.02 \\
CXOKMZ J033827.5-352604  & 42  & 265  &  2.99E38  &  23.1$\pm$0.01 &  20.8$\pm$0.01 &  2.3$\pm$0.02 \\
CXOKMZ J033828.9-352602  & 43  & 264  &  4.88E38  & 25.43$\pm$0.07 & 23.25$\pm$0.07 & 2.18$\pm$0.1 \\
CXOKMZ J033827.2-352600  & 44  & 301  &  3.74E38  & 22.68$\pm$0.01 &  20.4$\pm$0.01 & 2.29$\pm$0.01 \\
CXOKMZ J033831.8-352603  & 46  & 320  &  2.46E39  & 23.23$\pm$0.01 & 20.97$\pm$0.01 & 2.25$\pm$0.02 \\
CXOKMZ J033831.6-352559  & 50  & 340  &  3.92E38  & 23.09$\pm$0.01 & 20.73$\pm$0.01 & 2.36$\pm$0.02 \\
CXOKMZ J033828.2-352551  & 51  & 351  &  1.14E38  & 24.66$\pm$0.04 & 22.33$\pm$0.03 & 2.33$\pm$0.05 \\
CXOKMZ J033834.9-352654  & 54  & 379  &  6.53E37  & 23.64$\pm$0.02 &  21.3$\pm$0.01 & 2.34$\pm$0.02 \\
CXOKMZ J033833.1-352553  & 59  & 453  &  3.40E38  & 24.05$\pm$0.02 & 21.69$\pm$0.02 & 2.36$\pm$0.03 \\
CXOKMZ J033836.3-352708  & 62  & 491  &  3.50E38  & 23.15$\pm$0.01 & 20.86$\pm$0.01 & 2.29$\pm$0.02 \\
CXOKMZ J033830.2-352507  & 81  & 549  &  1.75E38  &   23.$\pm$0.01 & 20.62$\pm$0.01 & 2.38$\pm$0.01 \\
\tableline
\enddata
\end{deluxetable}


\begin{deluxetable}{llllll}
\tablewidth{0pt}
\tablenum{5}
\tablecaption{NGC 3115 X-ray Binaries}
\startdata
    \\
\tableline
\tableline
	  No.& Chandra ID		    &	RA(J2000)  & Dec(J2000)  & L$_X$ (ergs s$^{-1})$   & HST-FOV  \\
\tableline
  1\tablenotemark{1} &       CXOKMZ J100513.9-074307 & 10:05:13.91 & -7:43:07.4 & 4.26E38 &  Y \\
  2 &       CXOKMZ J100514.2-074311 & 10:05:14.21 & -7:43:11.6 & 5.23E36 &  Y \\
  3 &       CXOKMZ J100514.2-074303 & 10:05:14.27 & -7:43:03.6 & 6.74E37 &  Y \\
  4 &       CXOKMZ J100513.7-074301 & 10:05:13.77 & -7:43:01.2 & 1.30E38 &  Y \\
\tableline
\enddata
\tablecomments{The complete version of this table is available in the electronic version of the Journal}
\end{deluxetable}

\begin{deluxetable}{rrrrrr}
\tablewidth{0pt}
\tablenum{6}
\tablecaption{NGC 3115 Globular Clusters}
\startdata
     \\
\tableline
\tableline
	  No.& RA(J2000)	    & Dec(J2000)  & V (mag)  & I (mag)  & V-I (mag)   \\
\tableline
 1 & 10:05:14.26 & -7:43:03.9 & 20.05$\pm$0.03 & 18.92$\pm$0.03 & 1.13$\pm$0.04 \\
 2 & 10:05:13.95 & -7:43:14.0 &  22.9$\pm$0.14 & 22.01$\pm$0.21 & 0.89$\pm$0.25 \\
 3 & 10:05:14.02 & -7:42:59.6 &   22.$\pm$0.07 & 21.04$\pm$0.08 & 0.96$\pm$0.11 \\
 4 & 10:05:14.46 & -7:43:05.5 & 19.71$\pm$0.01 &  18.8$\pm$0.01 & 0.91$\pm$0.02 \\
\tableline
\enddata
\tablecomments{The complete version of this table is available in the electronic version of the Journal}
\end{deluxetable}

\begin{deluxetable}{lllllll}
\tablewidth{0pt}
\tablenum{7}
\tablecaption{NGC 3115 GC-LMXBs}
\startdata
    \\
\tableline
\tableline
 Chandra ID		    &	XID  & OID  & L$_X$ (ergs s$^{-1})$   & V (mags) & I (mags) & V-I (mags)  \\
\tableline
CXOKMZ J100514.2-074303 &  3 &   1 & 6.74E37 & 20.05$\pm$0.03 & 18.92$\pm$0.03 & 1.13$\pm$0.04 \\
CXOKMZ J100514.5-074318 & 10 &  10 & 7.80E37 &  21.9$\pm$0.02 & 20.72$\pm$0.02 & 1.18$\pm$0.03 \\
CXOKMZ J100515.1-074252 & 18 &  18 & 3.20E37 & 20.99$\pm$0.05 & 19.82$\pm$0.05 & 1.17$\pm$0.07 \\
CXOKMZ J100513.3-074337 & 22 &  29 & 1.87E38 & 23.12$\pm$0.06 & 21.89$\pm$0.06 & 1.23$\pm$0.09 \\
CXOKMZ J100513.0-074338 & 24 &  34 & 9.54E36 & 21.55$\pm$0.02 & 20.44$\pm$0.02 & 1.11$\pm$0.03 \\
CXOKMZ J100516.2-074235 & 29 &  52 & 2.33E37 & 20.04$\pm$0.01 & 19.07$\pm$0.01 & 0.97$\pm$0.01 \\
CXOKMZ J100517.1-074319 & 30 &  57 & 4.54E37 & 20.52$\pm$0.01 & 19.36$\pm$0.01 & 1.16$\pm$0.01 \\
CXOKMZ J100513.1-074217 & 31 &  63 & 3.11E37 & 19.97$\pm$0.01 &  18.9$\pm$0.01 & 1.07$\pm$0.01 \\
CXOKMZ J100518.3-074243 & 35 & 106 & 1.34E37 & 21.39$\pm$0.01 &  20.2$\pm$0.01 & 1.19$\pm$0.02 \\
\tableline
\enddata
\end{deluxetable}

\begin{deluxetable}{llllll}
\tablewidth{0pt}
\tablenum{8}
\tablecaption{NGC 3379 X-ray Binaries}
\startdata
    \\
\tableline
\tableline
	  No.& Chandra ID		    &	RA(J2000)  & Dec(J2000)  & L$_X$ (ergs s$^{-1})$   & HST-FOV  \\
\tableline
 1 & CXOKMZ J104749.7+123452 & 10:47:49.78 &  12:34:52.0 & 2.29E37 & Y \\
 2 & CXOKMZ J104749.8+123454 & 10:47:49.83 &  12:34:54.9 & 2.24E38 & Y \\
 3 & CXOKMZ J104749.6+123458 & 10:47:49.66 &  12:34:58.1 & 9.44E37 & Y \\
 4 & CXOKMZ J104749.4+123459 & 10:47:49.47 &  12:34:59.6 & 2.73E38 & Y \\
\tableline
\enddata
\tablecomments{The complete version of this table is available in the electronic version of the Journal}
\end{deluxetable}

\begin{deluxetable}{rrrrrr}
\tablewidth{0pt}
\tablenum{9}
\tablecaption{NGC 3379 Globular Clusters}
\startdata
     \\
\tableline
\tableline
	  No.& RA(J2000)	    & Dec(J2000)  & V (mag)  & I (mag)  & V-I (mag)   \\
\tableline
 1 & 10:47:49.95 & 12:34:53.1 & 21.04$\pm$0.04 & 19.88$\pm$0.04 & 1.16$\pm$0.06  \\
 2 & 10:47:50.05 & 12:34:55.5 & 22.25$\pm$0.08 &  21.2$\pm$0.09 & 1.06$\pm$0.12  \\
 3 & 10:47:50.18 & 12:34:55.1 &  21.5$\pm$0.03 & 20.65$\pm$0.04 & 0.86$\pm$0.05  \\
 4 & 10:47:49.75 & 12:35:05.7 & 22.94$\pm$0.05 &  21.9$\pm$0.07 & 1.05$\pm$0.08  \\
\tableline
\enddata
\tablecomments{The complete version of this table is available in the electronic version of the Journal}
\end{deluxetable}

\begin{deluxetable}{lllllll}
\tablewidth{0pt}
\tablenum{10}
\tablecaption{NGC 3379 GC-LMXBs}
\startdata
    \\
\tableline
\tableline
 Chandra ID		    &	XID  & OID  & L$_X$ (ergs s$^{-1})$   & V (mags) & I (mags) & V-I (mags)  \\
\tableline
CXOKMZ J104750.1+123455 &  9 &  3 & 2.83E38 & 21.5$\pm$0.03 & 20.65$\pm$0.04 & 0.86$\pm$0.05 \\
CXOKMZ J104750.3+123506 & 11 &  5 & 3.34E37 &21.45$\pm$0.01 & 20.28$\pm$0.01 & 1.18$\pm$0.02 \\
CXOKMZ J104750.4+123436 & 13 &  7 & 2.76E37 &20.94$\pm$0.01 & 19.82$\pm$0.01 & 1.12$\pm$0.02 \\
CXOKMZ J104752.7+123508 & 24 & 32 & 2.34E38 &21.08$\pm$0.01 & 20.24$\pm$0.01 & 0.84$\pm$0.01 \\
CXOKMZ J104751.0+123549 & 30 & 39 & 2.49E37 &19.99$\pm$  0. & 18.95$\pm$  0. & 1.04$\pm$0.01 \\
CXOKMZ J104754.2+123529 & 33 & 52 & 1.50E37 &23.15$\pm$0.03 & 21.93$\pm$0.03 & 1.22$\pm$0.04 \\
CXOKMZ J104752.6+123337 & 37 & 57 & 7.43E38 &21.82$\pm$0.01 & 20.69$\pm$0.01 & 1.12$\pm$0.02 \\
\tableline
\enddata
\end{deluxetable}

\begin{deluxetable}{llllll}
\tablewidth{0pt}
\tablenum{11}
\tablecaption{NGC 4594 X-ray Binaries}
\startdata
    \\
\tableline
\tableline
	  No.& Chandra ID		    &	RA(J2000)  & Dec(J2000)  & L$_X$ (ergs s$^{-1})$   & HST-FOV  \\
\tableline
  1\tablenotemark{1} & CXOKMZ J123959.4-113722 & 12:39:59.48 & -11:37:23.0 & 1.35E40 &  Y \\
  2 & CXOKMZ J123959.4-113727 & 12:39:59.45 & -11:37:27.1 & 3.78E38 &  Y \\
  3 & CXOKMZ J123959.1-113719 & 12:39:59.11 & -11:37:19.8 & 1.79E38 &  Y \\
  4 & CXOKMZ J123959.7-113716 & 12:39:59.78 & -11:37:16.6 & 1.72E38 &  Y \\
\tableline
\enddata
\tablecomments{The complete version of this table is available in the electronic version of the Journal}
\end{deluxetable}

\begin{deluxetable}{rrrrrr}
\tablewidth{0pt}
\tablenum{12}
\tablecaption{NGC 4594 Globular Clusters}
\startdata
     \\
\tableline
\tableline
	  No.& RA(J2000)	    & Dec(J2000)  & V (mag)  & I (mag)  & V-I (mag)   \\
\tableline
    1 & 12:39:59.47 & -11:37:27.3 & 21.14$\pm$0.06 & 19.92$\pm$0.06 & 1.23$\pm$0.09 \\
    2 & 12:39:59.65 & -11:37:18.8 & 21.44$\pm$0.06 & 20.39$\pm$0.07 & 1.06$\pm$0.1 \\
    3 & 12:39:59.08 & -11:37:19.8 & 20.43$\pm$0.02 & 19.33$\pm$0.03 &  1.1$\pm$0.03 \\
    4 & 12:39:59.69 & -11:37:29.2 & 22.34$\pm$0.1 & 21.17$\pm$0.09 & 1.17$\pm$0.13 \\
\tableline
\enddata
\tablecomments{The complete version of this table is available in the electronic version of the Journal}
\end{deluxetable}

\begin{deluxetable}{lllllll}
\tablewidth{0pt}
\tablenum{13}
\tablecaption{NGC 4594 GC-LMXBs}
\startdata
    \\
\tableline
\tableline
 Chandra ID		    &	XID  & OID  & L$_X$ (ergs s$^{-1})$   & V (mags) & I (mags) & V-I (mags)  \\
\tableline
CXOKMZ J123959.4-113727 &  2 &    1 &  3.78E38 & 21.14$\pm$0.06 & 19.92$\pm$0.06 & 1.23$\pm$0.09 \\
CXOKMZ J123959.1-113719 &  3 &    3 &  1.79E38 & 20.43$\pm$0.02 & 19.33$\pm$0.03 &  1.1$\pm$0.03 \\
CXOKMZ J124000.3-113723 &  9 &   17 &  3.10E38 &  19.4$\pm$0.01 & 18.21$\pm$0.01 & 1.19$\pm$0.01 \\
CXOKMZ J124001.0-113708 & 26 &   34 &  6.28E37 & 19.83$\pm$0.01 & 18.72$\pm$0.01 & 1.11$\pm$0.01 \\
CXOKMZ J124000.9-113702 & 29 &   39 &  5.15E37 & 22.49$\pm$0.07 & 21.22$\pm$0.05 & 1.27$\pm$0.08 \\
CXOKMZ J124002.1-113723 & 37 &   57 &  3.70E37 &   21.$\pm$0.03 & 19.76$\pm$0.02 & 1.24$\pm$0.04 \\
CXOKMZ J124002.0-113707 & 38 &   62 &  6.31E37 & 21.68$\pm$0.03 & 20.51$\pm$0.03 & 1.17$\pm$0.04 \\
CXOKMZ J124002.2-113801 & 49 &   98 &  6.06E37 & 19.66$\pm$0.01 & 18.46$\pm$0.01 &  1.2$\pm$0.01 \\
CXOKMZ J123959.3-113828 & 57 &  119 &  1.87E38 & 20.36$\pm$0.01 & 19.17$\pm$0.01 & 1.19$\pm$0.01 \\
CXOKMZ J124003.1-113645 & 59 &  122 &  2.82E37 &  22.7$\pm$0.06 & 21.66$\pm$0.04 & 1.03$\pm$0.07 \\
CXOKMZ J123958.9-113838 & 64 &  146 &  1.25E38 & 22.15$\pm$0.03 & 21.01$\pm$0.02 & 1.14$\pm$0.03 \\
CXOKMZ J123959.3-113846 & 70 &  156 &  2.78E37 & 21.21$\pm$0.02 & 20.05$\pm$0.01 & 1.16$\pm$0.02 \\
CXOKMZ J124005.7-113711 & 76 &  171 &  1.91E38 &  22.7$\pm$0.1 & 21.42$\pm$0.07 & 1.28$\pm$0.12 \\
CXOKMZ J124004.6-113829 & 80 &  185 &  3.90E37 & 21.57$\pm$0.02 &  20.4$\pm$0.01 & 1.17$\pm$0.03 \\
CXOKMZ J124007.0-113753 & 86 &  191 &  3.68E38 & 21.01$\pm$0.01 & 19.82$\pm$0.01 &  1.2$\pm$0.02 \\
\tableline
\enddata
\end{deluxetable}

\begin{deluxetable}{llllll}
\tablewidth{0pt}
\tablenum{14}
\tablecaption{NGC 4649 X-ray Binaries}
\startdata
    \\
\tableline
\tableline
	  No.& Chandra ID		    &	RA(J2000)  & Dec(J2000)  & L$_X$ (ergs s$^{-1})$   & HST-FOV  \\
\tableline
  1\tablenotemark{1} & CXOKMZ J124339.9+113310 & 12:43:39.98 & 11:33:10.0 & 1.79E39 &  Y \\
  2\tablenotemark{2} & CXOKMZ J124339.2+113313 & 12:43:39.21 & 11:33:13.7 & 1.06E38 &  Y \\
  3 & CXOKMZ J124340.2+113324 & 12:43:40.25 & 11:33:24.2 & 1.95E38 &  Y \\
  4 & CXOKMZ J124340.0+113251 & 12:43:40.03 & 11:32:51.8 & 9.06E37 &  N \\
\tableline
\enddata
\tablecomments{The complete version of this table is available in the electronic version of the Journal}
\end{deluxetable}

\begin{deluxetable}{rrrrrr}
\tablewidth{0pt}
\tablenum{15}
\tablecaption{NGC 4649 Globular Clusters}
\startdata
     \\
\tableline
\tableline
	  No.& RA(J2000)	    & Dec(J2000)  & V (mag)  & I (mag)  & V-I (mag)   \\
\tableline
   1\tablenotemark{1} & 12:43:39.98 & 11:33:08.4 & 21.84$\pm$0.11 & 20.64$\pm$0.14 & 1.2$\pm$0.18 \\
   2\tablenotemark{1} & 12:43:39.92 & 11:33:08.2 &  22.2$\pm$0.14 & 20.87$\pm$0.15 &1.33$\pm$0.2 \\
   3\tablenotemark{1} & 12:43:39.86 & 11:33:14.5 & 21.03$\pm$0.03 & 20.01$\pm$0.04 &1.02$\pm$0.04 \\
   4\tablenotemark{1} & 12:43:40.04 & 11:33:14.8 & 22.51$\pm$0.1 &  21.2$\pm$0.1 &1.31$\pm$0.14 \\
\tableline
\enddata
\tablecomments{The complete version of this table is available in the electronic version of the Journal}
\end{deluxetable}

\begin{deluxetable}{lllllll}
\tablewidth{0pt}
\tablenum{16}
\tablecaption{NGC 4649 GC-LMXBs}
\startdata
    \\
\tableline
\tableline
 Chandra ID		    &	XID  & OID  & L$_X$ (ergs s$^{-1})$   & V (mags) & I (mags) & V-I (mags)  \\
\tableline
CXOKMZ J124337.8+113327 &  16 & 101 & 3.30E38 &  21.21$\pm$0.01 & 20.1$\pm$0.01 & 1.12$\pm$0.01 \\
CXOKMZ J124337.4+113309 &  18 & 109 & 1.51E38 &  21.01$\pm$0.01 &19.86$\pm$0.01 & 1.15$\pm$0.01 \\
CXOKMZ J124338.7+113342 &  20 & 112 & 1.26E38 &  22.13$\pm$0.01 &20.91$\pm$0.01 & 1.23$\pm$0.02 \\
CXOKMZ J124338.2+113343 &  23 & 136 & 6.81E37 &  22.08$\pm$0.01 &20.85$\pm$0.01 & 1.23$\pm$0.02 \\
CXOKMZ J124341.6+113351 &  26 & 167 & 1.92E38 &  21.78$\pm$0.01 &20.56$\pm$0.01 & 1.22$\pm$0.01 \\
CXOKMZ J124342.6+113340 &  27 & 176 & 2.36E37 &  21.75$\pm$0.01 &20.72$\pm$0.01 & 1.03$\pm$0.01 \\
CXOKMZ J124336.2+113312 &  34 & 205 & 1.48E38 &  22.16$\pm$0.01 &20.94$\pm$0.01 & 1.22$\pm$0.02 \\
CXOKMZ J124343.3+113341 &  36 & 225 & 1.05E38 &   23.4$\pm$0.03 &22.18$\pm$0.03 & 1.22$\pm$0.04 \\
CXOKMZ J124336.7+113348 &  37 & 236 & 1.04E38 &  22.22$\pm$0.01 &20.94$\pm$0.01 & 1.28$\pm$0.02 \\
CXOKMZ J124338.1+113404 &  38 & 237 & 1.15E38 &  22.81$\pm$0.02 &21.63$\pm$0.02 & 1.18$\pm$0.03 \\
CXOKMZ J124342.9+113355 &  40 & 250 & 1.04E38 &  21.75$\pm$0.01 &20.53$\pm$0.01 & 1.22$\pm$0.01 \\
CXOKMZ J124335.8+113350 &  44 & 300 & 8.20E37 &  22.24$\pm$0.01 &20.92$\pm$0.01 & 1.32$\pm$0.02 \\
CXOKMZ J124341.5+113428 &  48 & 348 & 8.06E37 &  20.97$\pm$0.01 & 19.8$\pm$0.01 & 1.17$\pm$0.01 \\
CXOKMZ J124334.3+113310 &  49 & 347 & 5.82E37 &  22.42$\pm$0.01 &21.36$\pm$0.01 & 1.06$\pm$0.02 \\
CXOKMZ J124334.6+113237 &  52 & 361 & 1.12E38 &  20.88$\pm$0.01 &19.67$\pm$0.01 & 1.21$\pm$0.01 \\
CXOKMZ J124343.0+113428 &  63 & 394 & 4.25E37 &  23.23$\pm$0.02 & 21.9$\pm$0.02 & 1.33$\pm$0.03 \\
\tableline
\enddata
\end{deluxetable}

\begin{deluxetable}{llllllll}
\tablenum{17}
\tablewidth{0pt}
\tablenum{17}
\tablecaption{Discriminant Analysis Weights }
\startdata
    \\
\tableline
\tableline
Galaxy		&	Luminosity	&  Color\tablenotemark{1}	& Distance 	& Random1\tablenotemark{2}	& Random2\tablenotemark{2}	& p-value\tablenotemark{3} 	& \% correct\tablenotemark{3} \\
NGC 1399	& 0.704(0.773)	& -0.561(-0.610)	& 0.077(0.182)	& -0.227(-0.221)	& 0.211(0.237)	&	0.000	& 70\% \\
NGC 1399	& 0.757(0.812)	& -0.580(-0.641)	& 0.071(0.191)	& {\nodata}	& {\nodata}	&	0.000	& 69\% \\
NGC 3115	& 0.756(0.816)	& -0.399(-0.414)	& 0.403(0.482)	& -0.003(-0.052)	& 0.175(0.135)	&	0.006	& 74\% \\
NGC 3115	& 0.776(0.829)	& -0.406(-0.420)	& 0.380(0.489)	& {\nodata}	& {\nodata}	&	0.001	& 72\% \\
NGC 3379	& 0.978(0.959)	& -0.070(0.204) 	& 0.086(0.217)	& 0.131(0.161)	& 0.233(0.158)	&	0.192	& 74\% \\
NGC 3379	& 0.990(0.819)	& -0.017(0.212) 	& 0.071(0.225)	& {\nodata}	& {\nodata}	&	0.072	& 74\% \\
NGC 4594	& 0.711(0.746)	& -0.614(-0.664)	& -0.001(0.048)	& -0.086(-0.042)	& 0.216(0.273)	&	0.000	& 76\% \\
NGC 4594	& 0.732(0.767)	& -0.643(-0.682)	& -0.002(0.049)	& {\nodata}	& {\nodata}	&	0.000	& 79\% \\
NGC 4649	& 0.911(0.727)	& -0.467(-0.404)	& -0.372(-0.091)	& 0.242(0.171)	& 0.296(0.249)	&	0.000	& 79\% \\
NGC 4649	& 0.940(0.787)	& -0.512(-0.438)	& -0.364(0.098)	& {\nodata}	& {\nodata}	&	0.000	& 79\% \\
\multicolumn{8}{c}{\underline{BRIGHT GLOBULAR CLUSTERS\tablenotemark{4}}}\\
NGC 1399	& 0.583(0.660)	& -0.632(-0.661)	& 0.130(0.108)	& 0.353(0.372)	& -0.230(-0.225)&	0.000	& 70\% \\
NGC 1399	& 0.680(0.725)	& -0.677(-0.727)	& 0.125(0.119)	& {\nodata}	& {\nodata}	&	0.000	& 67\% \\
NGC 3115	& 0.823(0.811)	& -0.348(-0.385)	& 0.480(0.390)	& -0.121(-0.104)	& 0.040(-0.010)	&	0.023	& 74\% \\
NGC 3115	& 0.839(0.818)	& -0.339(-0.388)	& 0.464(0.393)	& {\nodata}	& {\nodata}	&	0.005	& 75\% \\
NGC 3379	& 0.861(0.863)	& 0.263(0.362)  	& 0.326(0.257) 	& 0.162(0.101)	& 0.384(0.160)	&	0.335	& 79\% \\
NGC 3379	& 0.884(0.939)	& 0.288(0.394)  	& 0.205(0.280) 	& {\nodata}	& {\nodata}	&	0.167	& 76\% \\
NGC 4594	& 0.696(0.545)	& -0.833(-0.698)	& -0.012(0.062)	& -0.024(-0.087)	& 0.241(0.158)	&	0.000	& 79\% \\
NGC 4594	& 0.707(0.562)	& -0.840(-0.719)	& -0.019(-0.064)	& {\nodata}	& {\nodata}	&	0.000	& 80\% \\
NGC 4649	& 0.756(0.619)	& -0.579(-0.492)	& -0.368(-0.223)& 0.279(0.218)	& 0.342(0.304)	&	0.000	& 77\% \\
NGC 4649	& 0.801(0.691)	& -0.652(-0.549)	& -0.356(-0.249)	& {\nodata}	& {\nodata}	&	0.000	& 78\% \\
\tableline		 
\enddata
\tablenotetext{1} {B-I for NGC 1399 and V-I for the other galaxies.}
\tablenotetext{2} {Random1 and Random2 are dummy Normal and Poisson random variables respectively. Note that the two sets of numbers listed for each galaxy present the results with and without the random variables in order to help the reader judge which of the variables have significant power. Since small correlations between the random variables and the discriminating variables may affect computed discriminating power of the latter, the tests without random variables should be used to gauge the relative discriminating power in the various physical quantities.} 
\tablenotetext{3} {The ``p-value" denotes the significance of the null hypothesis that there is no discriminating power in the variables based on Wilks' lambda statistic, while ``\% correct" reports the percentage of cases classified in the right group by DA.} 
\tablenotetext{4} {Clusters brighter than the turnover magnitude of globular cluster luminosity function.}
\end{deluxetable}

\end{document}